\ifdefined\pdfminorversion\pdfminorversion=7\fi
\ifdefined\pdfsuppresswarningpagegroup\pdfsuppresswarningpagegroup=1\fi

\documentclass[
  aps,
  prd,
  reprint,
  onecolumn,
  groupedaddress,
  longbibliography
]{revtex4-2}

\usepackage{amsmath,amssymb}
\usepackage{bm}
\usepackage{graphicx}
\usepackage{dcolumn}
\usepackage{multirow}
\usepackage{booktabs}
\usepackage{float}
\usepackage[hidelinks]{hyperref}
\usepackage{url}

\begin{document}
\raggedbottom
\clubpenalty=10000
\widowpenalty=10000
\displaywidowpenalty=10000
\setlength{\textfloatsep}{10pt plus 2pt minus 2pt}
\setlength{\floatsep}{8pt plus 2pt minus 2pt}
\setlength{\intextsep}{10pt plus 2pt minus 2pt}
\renewcommand{\topfraction}{0.95}
\renewcommand{\textfraction}{0.05}
\renewcommand{\floatpagefraction}{0.8}

\title{Real-Time Doppler-Frequency Measurement and Frequency-Pulling Suppression in Arm-locking}

\author{Zheng-Rong Xiang$^{1,2}$}
\author{Jun Ke$^{3}$}\email[E-mail: ]{junke@hust.edu.cn}
\author{Tian-Chi Zhu$^{1}$}
\author{Hao-Xin Xu$^{1}$}
\author{Cheng-Gang Shao$^{3}$}
\author{Jie Luo$^{1}$}\email[E-mail: ]{luojiethanks@126.com }

\affiliation{$^{1}$School of Mechanical Engineering and Electronic Information, China University of Geosciences, 430074, Wuhan, China}
\affiliation{$^{2}$School of Physics and Information Engineering, Guangxi Science $\&$ Technology Normal University, 546199, Laibin, China}
\affiliation{$^{3}$National Gravitation Laboratory, MOE Key Laboratory of Fundamental Physical Quantities Measurement, and School of Physics, Huazhong University of Science and Technology, 430074, Wuhan, China}

\begin{abstract}
Arm-locking is an important technique for suppressing laser frequency noise in space-based gravitational-wave detection. However, stronger noise suppression inevitably increases the magnitude of frequency pulling, thereby prolonging the lock-acquisition time and increasing the risk of system instability. To address this issue, this paper proposes a real-time Doppler-frequency measurement and frequency-pulling suppression method based on a linear-quadratic-gaussian controller. To achieve a balance between noise suppression and pulling suppression, a time-varying control strategy is adopted. At the initial stage of the time-varying control process, the primary objective is to rapidly limit frequency pulling. Subsequently, the level of pulling suppression is gradually relaxed while the capability for laser frequency-noise suppression is progressively enhanced, so that the residual frequency noise satisfies the time-delay interferometry (TDI) requirement by the end of the transition. Comprehensive performance simulations are conducted for several arm-locking configurations. The results show that the proposed suppressor can reduce frequency pulling within approximately 1000 s while limiting its magnitude to below 100 MHz, thereby eliminating the need for the conventional Doppler-frequency pre-estimation stage. When a Fabry–Perot laser is used, the amplitude spectral density (ASD) of the closed-loop laser frequency noise meets the requirements of second-generation and first-generation TDI after 2000 s and 9200 s of closed-loop operation, respectively. These results demonstrate that the proposed suppressor provides both rapid frequency-pulling suppression and a high level of laser frequency-noise suppression. This new method provide a feasible way for rapid acquisition and locking of arm-locking.
\end{abstract}
\maketitle

\section{Introduction}

Space-based gravitational-wave (GW) observatories such as LISA \cite{amaroSeoane2017}, Taiji \cite{luo2021Taiji}, and TianQin \cite{luo2016TianQin} are designed to probe low-frequency gravitational waves that are inaccessible to ground-based detectors. A passing GW changes the relative distances between the test masses, and these variations can be measured by exchanging coherent laser beams between widely separated spacecraft. A major limitation of such measurements is laser frequency noise. Since the interferometric arms are unequal and time dependent, laser frequency noise does not cancel directly in the interferometric measurements, and thus overwhelm the GW signals \cite{tintoArmstrong1999}. Therefore, laser-frequency pre-stabilization, arm locking and time-delay interferometry (TDI) techniques are used to suppress the residual laser frequency noise \cite{tintoArmstrong1999,tintoDhurandhar2021}. Arm locking provides an additional stage of laser-frequency stabilization by using the long interferometer arms themselves as frequency references. It can therefore complement optical-cavity pre-stabilization and reduce the level of laser frequency noise entering the subsequent TDI processing. In practice, the arm-length references are neither static nor instantaneous. The Doppler shifts caused by spacecraft motion, introduce nontrivial dynamics into the feedback loop, making closed-loop stability and controlled lock acquisition essential for realizing the full noise-suppression capability of arm locking.

Arm-locking was first proposed by Sheard et al. \cite{sheard2003} and has since progressed through numerical studies, electronic-delay experiments, and laboratory demonstrations \cite{sylvestre2004,thorpe2005,sheard2005,herz2005}. Studies of constellation dynamics \cite{dhurandhar2005}, hardware models \cite{thorpe2006}, and tunable cavity prestabilization \cite{thorpe2008} established the basis for more realistic implementations. Dual-arm-locking \cite{sutton2008} addressed the sensor nulls that limit single-arm control by combining measurements from unequal arms, while tunable prestabilized lasers supported frequency control and lock acquisition \cite{livas2009}. This improved sensor response comes with a practical difficulty: errors in the estimated interspacecraft Doppler frequencies drive laser-frequency pulling. McKenzie et al. \cite{mckenzie2009} characterized this coupling and introduced modified dual-arm-locking to mitigate pulling and low-frequency noise coupling. Subsequent experiments and simulations examined Doppler-frequency errors, long delays, and mission-specific operating conditions \cite{wand2009,yu2010,yu2011Experiment,yu2011Thesis,thorpe2011,yu2014,thorpe2016}. Together, these studies established the feasibility of arm-locking while exposing a persistent tension between strong noise suppression and tolerable frequency excursions.

More recent work has refined both the sensing architecture and its implementation. For the updated LISA design, Ghosh et al. \cite{ghosh2022} quantified Doppler-induced pulling and the averaging needed to limit it. Frequency-comb-assisted sensing \cite{wu2022}, combined cavity-and-arm control \cite{valliyakalayil2022}, and developments in frequency setting, common-arm sensing, and transfer-oscillator techniques \cite{zhang2023,ke2023,xu2024} have expanded the available stabilization strategies; combined cavity-and-arm-locking has also been demonstrated experimentally \cite{valliyakalayil2024}. Meanwhile, studies of clock-jitter and laser phase noise \cite{wu2024} and controller transients \cite{zhang2024} have clarified performance and stability constraints, including the limitations of phase-margin criteria for delayed feedback. Mission simulators \cite{li2025}, catalogs of laser-locking configurations \cite{heinzel2025}, dual-reference stabilization \cite{valliyakalayil2025}, and spacecraft controller designs \cite{su2025,zhang2025} provide further tools for evaluating practical implementations. Taken together, these advances sharpen the remaining acquisition problem: Doppler-frequency estimation errors must be brought under control quickly, while preserving the steady-state noise suppression that motivates arm-locking.

Despite these advances in steady-state noise suppression, rapid convergence during lock acquisition remains a fundamental challenge. Conventional approaches reduce Doppler-frequency estimation error by long-time averaging of interspacecraft phase measurements or ranging data. Although such long-integration strategies lower the pulling rate, they also increase the waiting time before gravitational-wave observations can begin. Because the Doppler-frequency varies in time, a long-time average also lags the instantaneous value, and the benefit of further increasing the averaging time eventually becomes limited. More importantly, laser-frequency-noise suppression and frequency pulling are intrinsically in conflict: increasing the low-frequency loop gain improves noise rejection but simultaneously amplifies the system response to Doppler-frequency estimation error, producing larger pulling and longer convergence. Active suppression of frequency pulling without sacrificing steady-state noise performance is therefore a key problem for arm-locking.

To address this problem, we propose a real-time Doppler-frequency measurement and frequency-pulling suppression method based on a linear-quadratic-Gaussian (LQG) controller. The LQG suppressor combines a Kalman filter (KF) and a linear-quadratic regulator (LQR): the KF estimates the Doppler-frequency in real-time during closed-loop arm-locking, while the LQR uses the estimated state to generate an optimal suppression command that actively limits frequency-pulling. Unlike conventional schemes that perform a long Doppler pre-frequency estimation before loop closure, the proposed method embeds Doppler-frequency estimation directly into the arm-lock acquisition stage so that measurement and suppression proceed simultaneously. To balance pulling suppression during acquisition against laser-frequency-noise rejection in steady-stage, we further introduce a time-varying LQR strategy. Large weights are assigned to pulling-related states immediately after loop closure to prevent large excursions from developing. After the Doppler-frequency has been estimated and the pulling is controlled, the pulling-suppression weights are gradually reduced, allowing the arm-locking loop to recover high-gain laser-noise suppression.

The main contributions of this work are threefold. First, we propose a method for real-time Doppler frequency measurement and suppression in arm-locking systems, eliminating the need for a dedicated Doppler frequency pre-estimation stage before loop closure. Second, we design an LQG-based suppressor to achieve rapid suppression of Doppler-induced frequency pulling. Third, we develop a time-varying weighting strategy to balance frequency noise suppression and frequency-pulling suppression, with the aim of minimizing the lock acquisition time.

The remainder of this paper is organized as follows. Section II analyzes the mechanism by which Doppler-frequency errors induce frequency-pulling and establishes measurement and suppression models. Section III develops the state-space model of the LQG suppressor and presents the KF estimator, LQR regulator, and time-varying LQG parameter strategy. Section IV validates the proposed approach through numerical simulations, with emphasis on pulling amplitude and convergence time. Section V discusses the real-time measurement and pulling-suppression performance, and Sec. VI concludes the paper.

\section{Modeling}

\subsection{Arm-locking model}

\begin{figure}[H]
\centering
\begin{minipage}[t]{0.48\textwidth}
\centering
\includegraphics[width=\linewidth]{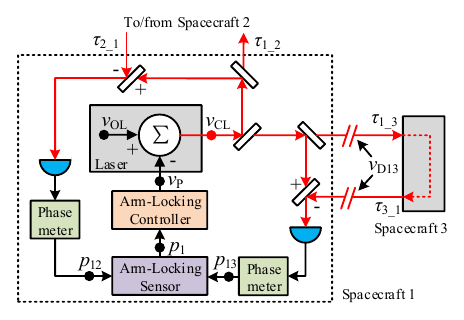}
\par\smallskip (a)
\end{minipage}
\hfill
\begin{minipage}[t]{0.50\textwidth}
\centering
\includegraphics[width=\linewidth]{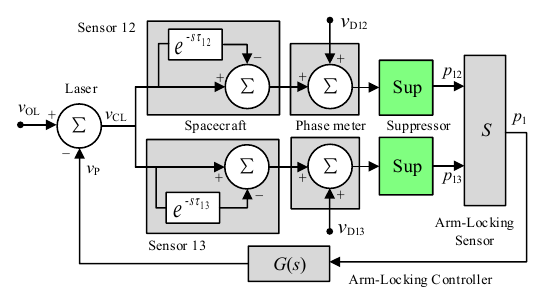}
\par\smallskip (b)
\end{minipage}
\caption{Arm-locking: (a) Physical configuration and (b) Model representation.}
\label{fig:1}
\end{figure}

The simplified arm-locking is shown in Fig.~\ref{fig:1}. Fig~\ref{fig:1}(a) gives the physical configuration, which consists of three spacecraft. Spacecraft 1 is the central spacecraft and contains the laser, phasemeter, arm-locking sensor, arm-locking controller, and other optical components. Spacecraft 2 (omitted for clarity) and spacecraft 3 are remote spacecraft that act as laser transponders in this study; their internal details are therefore omitted and only the transponder function is indicated by a red dashed line. Each remote spacecraft and the central spacecraft form an interferometer arm used as a gravitational-wave sensor. Red arrows denotes optical propagation, whereas black arrows denote electrical signal paths. The round-trip light-travel time between spacecraft 1 and spacecraft $x$($x$=2,3) is denoted by $\tau_{1x}$ and equals the sum of the two one-way propagation times ($\tau_{1\_x}$ and $\tau_{x\_1}$). The phasemeter output, $p_{\mathrm{1x}}$, is processed by the arm-locking sensor to produce the arm-locking signal, $p_{1}$. Relative spacecraft motion generates the Doppler-frequency, $v_{\mathrm{D1x}}$, and hence the frequency-pulling, $v_{\mathrm{P}}$. The intrinsic laser frequency-noise is denoted by $v_{\mathrm{OL}}$ in the open loop and becomes the residual closed-loop noise, $v_{\mathrm{CL}}$, after arm-locking suppression. The primary purpose of arm-locking is to minimize $v_{\mathrm{CL}}$.

The model representation is shown in Fig.~\ref{fig:1}(b). The two green blocks are the suppressors developed in this work; their positions in the schematic differ from the practical implementation and are shown only to illustrate the principle. They suppress frequency-pulling caused by the Doppler-frequency. The laser emitter is modeled as a subtractor and the phasemeter as an adder. Each interferometer arm is modeled as a sensor consisting of a delay element and a subtractor, and the arm-locking controller is modeled by the transfer function $G(s)$. The arm-locking sensor is represented by the matrix $\bm{S}=[s_1\ s_2]$. The phase signals form the input vector, $p_{12}$ and $p_{13}$, is multiplied by $\bm{S}$ to obtain $p_{1}$:
\begin{equation*}
p_1=\bm{S}\begin{bmatrix}p_{12}\\p_{13}\end{bmatrix}=s_1p_{12}+s_2p_{13}
\tag{1}\label{eq:1}
\end{equation*}
Each arm-locking sensor configuration has a corresponding matrix $\bm{S}$, see Ref. \cite{mckenzie2009}.

\subsection{Propagation of the Doppler frequency in the arm-locking}

\begin{figure}[H]
\centering
\includegraphics[width=0.50\textwidth]{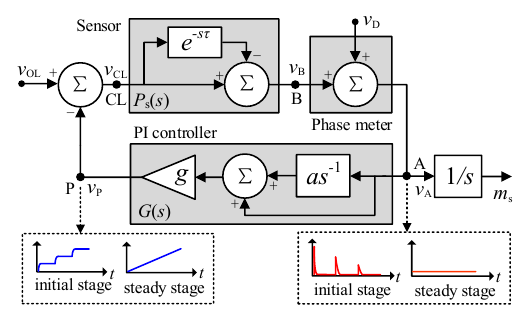}
\caption{Single-arm-locking with a PI controller. In the left subplot, the blue curve is the frequency-pulling, and in the right subplot, the red curve is the output signal of the phase meter.}
\label{fig:2}
\end{figure}

We use a single-arm system to analyze the propagation of Doppler frequency through the arm-locking loop. The model in Fig.~\ref{fig:2} consists of a single interferometric-arm sensor, a phasemeter, and a controller. In the Laplace domain, the arm-sensor transfer function is:
\begin{equation*}
P_{\mathrm s}(s)=1-e^{-s\tau}
\tag{2}\label{eq:2}
\end{equation*}
Here $s$ is the Laplace variable and $\tau$ is the round-trip light-travel time, approximately 16.6 s for new LISA. For analytical convenience, we use the simple PI arm-locking controller of Ref. \cite{zhang2024}, whose transfer function is:
\begin{equation*}
G(s)=g+g a s^{-1}=g\left(1+a s^{-1}\right)
\tag{3}\label{eq:3}
\end{equation*}
where $g$ denotes the controller gain and $a$ denotes the zero.

In Fig.~\ref{fig:2}, the Doppler-frequency, $v_{\mathrm{D}}$, contains a constant term, 1st- and 2nd-order time-derivative terms, and higher-order terms. For clarity, the first three are denoted by $v_{\mathrm{D0}}$, $v_{\mathrm{D1}}$, and $v_{\mathrm{D2}}$, respectively:
\begin{equation*}
v_{\mathrm D}(t)=v_{\mathrm{D0}}+\int_0^t v_{\mathrm{D1}}\,dt+\int_0^t\int_0^{y}v_{\mathrm{D2}}\,dx\,dy+\text{(higher order terms)}
\tag{4}\label{eq:4}
\end{equation*}
For the following discussion, assume temporarily that $v_{\mathrm{D0}}$ is the dominant component and that $v_{\mathrm{OL}}$=0. After loop closure, $v_{\mathrm{D}}$ enters the system as a step signal. It produces the controller input, $v_{\mathrm{A}}$, at point A, the frequency-pulling, $v_{\mathrm{P}}$, at point P, and the phasemeter input, $v_{\mathrm{B}}$, at point B. The integrator in the PI controller drives $v_{\mathrm{A}}$ exponentially toward zero, forming a narrow pulse as shown in the initial stage waveform in the subplot. Meanwhile, $v_{\mathrm{P}}$ approaches $v_{\mathrm{D}}$ exponentially and $v_{\mathrm{B}}$ approaches -$v_{\mathrm{D}}$; adding $v_{\mathrm{B}}$ and $v_{\mathrm{D}}$ in the phasemeter drives $v_{\mathrm{A}}$ toward zero, effectively blocking the continuous entry of $v_{\mathrm{D}}$ into the loop. At the same time, $v_{\mathrm{P}}$ propagates through the interferometer arm. After one delay $\tau$ it reaches point B, causing $v_{\mathrm{B}}$ to jump back to zero so that $v_{\mathrm{D}}$ reenters the loop and the next propagation cycle begins.

After the first propagation cycle, one pulse appears at point A and a pulling excursion of magnitude $v_{\mathrm{D}}$ appears at point P. Thereafter $v_{\mathrm{A}}$ becomes a pulse train with period, $\tau$, while $v_{\mathrm{P}}$ increases once per period. Because the integral term acts as a low-pass filter, the pulses are gradually smoothed and the change in $v_{\mathrm{P}}$ becomes slower. After several cycles, the pulse train approaches a DC signal and $v_{\mathrm{P}}$ approaches a ramp, marking the steady stage regime shown in the subplot.

\subsection{Doppler-frequency measurement model}

\subsubsection{Measurement of the single-arm-locking signal}

\begin{figure}[!htbp]
\centering
\begin{minipage}[t]{0.49\textwidth}
\centering
\includegraphics[width=\linewidth]{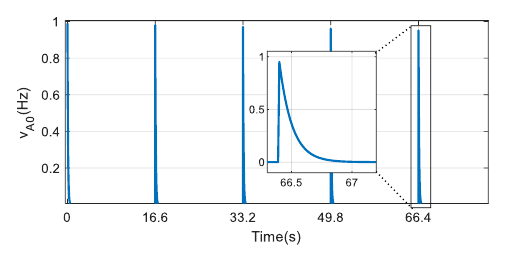}
\par\smallskip (a)
\end{minipage}
\hfill
\begin{minipage}[t]{0.49\textwidth}
\centering
\includegraphics[width=\linewidth]{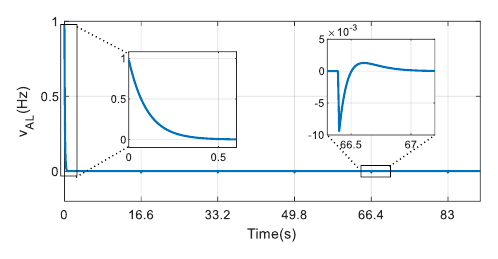}
\par\smallskip (b)
\end{minipage}
\caption{Time-domain curves of $v_{\mathrm{A0}}$ and $v_{\mathrm{AL}}$ for $v_{\mathrm{D0}}$=100 Hz, $g$=100, $a$=10, and $\tau$=16.6 s: (a) $v_{\mathrm{A0}}$ and (b) $v_{\mathrm{AL}}$.}
\label{fig:3}
\end{figure}

In the single-arm-locking, the quantity to be measured is $v_{\mathrm{D}}$, which contains $v_{\mathrm{D0}}$-$v_{\mathrm{D2}}$, and higher-order components. A suitable measurement point is required, because $v_{\mathrm{A}}$ is a periodic pulse signal whose magnitude is related to the Doppler-frequency, it is selected as the measurement signal. Taking $v_{\mathrm{D}}$ and $v_{\mathrm{A}}$ as the system input and output, respectively, gives:
\begin{equation*}
v_{\mathrm A}(s)=\frac{v_{\mathrm D}(s)}{1+P_{\mathrm s}(s)G(s)}
\tag{5}\label{eq:5}
\end{equation*}
We first consider the response produced by $v_{\mathrm{D0}}$. Because $v_{\mathrm{D0}}$ is a step input, it produces the step response, $v_{\mathrm{A0}}$, at point A. Using the derivation in Appendix A, it can be written as a sum of component terms:
\begin{equation*}
v_{\mathrm{A0}}(s)=\frac{v_{\mathrm{D0}}}{s+g(s+a)(1-e^{-s\tau})}=\sum_{n=1}^{\infty}v_{\mathrm{A0\_n}}(s)=v_{\mathrm{D0}}\sum_{n=1}^{\infty}\frac{\left[g(s+a)e^{-s\tau}\right]^{n-1}}{\left[(g+1)s+ga\right]^n}
\tag{6}\label{eq:6}
\end{equation*}
Here $v_{\mathrm{A0\_n}}$ denotes the nth term of $v_{\mathrm{A0}}$. The time-domain waveform obtained by inverse Laplace transformation is shown in Fig.~\ref{fig:3}(a), $v_{\mathrm{A0}}$ is a periodic sequence of narrow pulses. Because such narrow pulses are difficult to measure accurately, $v_{\mathrm{A0\_n}}$ is integrated over one interval of duration $\tau$ before measurement, and the integral is denoted by $I_{\mathrm{D0}}$. The pulse decays rapidly and is nearly zero by the end of the interval, so the final-value theorem can be used to approximate the integral:
\begin{equation*}
\lim_{t\to\infty}\int_0^t f(t)\,dt=\lim_{s\to0}f(s)
\tag{7}\label{eq:7}
\end{equation*}
Combining Eqs. (6) and (7) gives:
\begin{equation*}
I_{\mathrm{D0}}(n)=\int_{(n-1)\tau}^{n\tau}v_{\mathrm{A0\_n}}(t)\,dt\approx\lim_{s\to0}v_{\mathrm{A0\_n}}(s)=\frac{v_{\mathrm{D0}}}{ga},\quad(n\ge1)
\tag{8}\label{eq:8}
\end{equation*}
Eq.~\eqref{eq:8} shows that $I_{\mathrm{D0}}$ is proportional to $v_{\mathrm{D0}}$, so $v_{\mathrm{D0}}$ can be measured indirectly through $I_{\mathrm{D0}}$.

The open-loop laser noise, $v_{\mathrm{OL}}$, is another system input. Its contribution at point A is denoted by $v_{\mathrm{AL}}$ and is given by:
\begin{equation*}
v_{\mathrm{AL}}(s)=\frac{v_{\mathrm{OL}}P_{\mathrm s}(s)}{1+P_{\mathrm s}(s)G(s)}
 =\frac{v_{\mathrm{OL}}s(1-e^{-s\tau})}{s+g(s+a)(1-e^{-s\tau})}
\tag{9}\label{eq:9}
\end{equation*}
At loop closure, let $v_{\mathrm{L}}$ denote the instantaneous value of the laser noise. The noise enters the system as a step signal and produces the step response, $v_{\mathrm{AL}}$, at point A. Using Appendix B, $v_{\mathrm{AL}}$ can be written as a sum of component terms:
\begin{equation*}
v_{\mathrm{AL}}(s)=\frac{v_{\mathrm L}}{(g+1)s+ga}-v_{\mathrm L}\sum_{n=2}^{\infty}\frac{s\left[g(s+a)\right]^{n-2}e^{-(n-1)s\tau}}{\left[(g+1)s+ga\right]^n}
\tag{10}\label{eq:10}
\end{equation*}
Here $v_{\mathrm{AL\_n}}$ is the nth term. The inverse Laplace transform curve of $v_{\mathrm{AL}}$, shown in Fig.~\ref{fig:3}(b), gives a periodic pulse sequence. The first pulse has unit peak amplitude, whereas the largest peak of the later pulses is only about 0.01Hz.

Integrating $v_{\mathrm{AL}}$ piecewise over intervals of duration $\tau$ gives $I_{\mathrm{L}}$:
\begin{equation*}
I_{\mathrm L}(n)=\int_{(n-1)\tau}^{n\tau}v_{\mathrm{AL\_n}}(t)\,dt\approx\lim_{s\to0}v_{\mathrm{AL\_n}}(s)=\begin{cases}\dfrac{v_{\mathrm L}}{ga},&(n=1)\\[4pt]0,&(n>1)\end{cases}
\tag{11}\label{eq:11}
\end{equation*}
Eq.~\eqref{eq:11} shows that $I_{\mathrm{L}}$ is nonzero only during the 1st-cycle and vanishes in subsequent cycles.When the arm-locking loop is closed, $v_{\mathrm{OL}}$ and $v_{\mathrm{D}}$ enter simultaneously and the resulting $v_{\mathrm{A}}$ is:
\begin{equation*}
v_{\mathrm A}(n)=I_{\mathrm{D0}}(n)+I_{\mathrm L}(n)=\begin{cases}\dfrac{v_{\mathrm L}+v_{\mathrm{D0}}}{ga},&(n=1)\\[4pt]\dfrac{v_{\mathrm{D0}}}{ga},&(n>1)\end{cases}
\tag{12}\label{eq:12}
\end{equation*}
Because laser noise contain large low-frequency components, the instantaneous value $v_{\mathrm{L}}$ can be large at loop closure and make the 1st-cycle measurement inaccurate. To remove the influence of $v_{\mathrm{L}}$, the 1st-cycle measurement is discarded.

\begin{figure}[!htbp]
\centering
\includegraphics[width=0.50\textwidth]{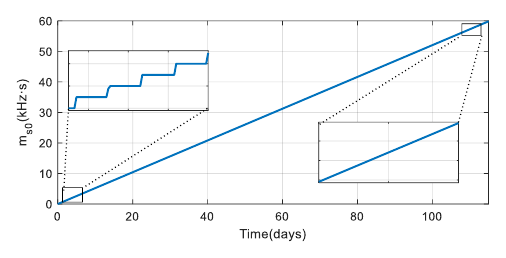}
\caption{Evolution of $m_{\mathrm{s0}}$ in the initial stage (left inset) and in the steady stage (right inset) regime after 112 days, for $v_{\mathrm{D0}}$=100 Hz.}
\label{fig:4}
\end{figure}

For practical implementation, $v_{\mathrm{A0}}$ is continuously integrated from the second cycle onward, and the resulting measurement is denoted by $m_{\mathrm{s0}}$. During the initial stage, each pulse of $v_{\mathrm{A0}}$ increases $m_{\mathrm{s0}}$ by $I_{\mathrm{D0}}$, so $m_{\mathrm{s0}}$ initially forms a monotonically increasing staircase. The integral action of the PI controller gradually smooths the staircase, and in steady stage $m_{\mathrm{s0}}$ becomes a ramp. Accordingly, the expression for $m_{\mathrm{s0}}$ is written separately for the initial and steady stage:
\begin{equation*}
m_{\mathrm{s0}}(t)=\int_0^t v_{\mathrm{A0}}(x)\,dx=\begin{cases}\dfrac{v_{\mathrm{D0}}}{ga}c,\quad c=\left\lfloor\dfrac{t}{\tau}\right\rfloor&\text{(initial stage)}\\[6pt]\dfrac{v_{\mathrm{D0}}}{ga\tau}t,&\text{(steady stage)}\end{cases}
\tag{13}\label{eq:13}
\end{equation*}
Here $\lfloor\,\cdot\,\rfloor$ is the floor function that returns the number($c$) of propagation cycles corresponding to time($t$). For example, if $t$=70 s and $\tau$=16.6 s, then $c$=4. The evolution of $m_{\mathrm{s0}}$ in the two stages is shown in Fig.~\ref{fig:4}.

\begin{figure}[!htbp]
\centering
\begin{minipage}[t]{0.49\textwidth}
\centering
\includegraphics[width=\linewidth]{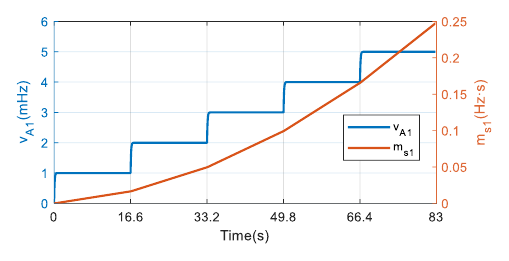}
\par\smallskip (a)
\end{minipage}
\hfill
\begin{minipage}[t]{0.49\textwidth}
\centering
\includegraphics[width=\linewidth]{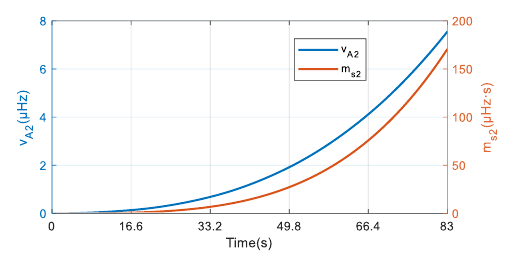}
\par\smallskip (b)
\end{minipage}
\caption{Signals generated by $v_{\mathrm{D1}}$ and $v_{\mathrm{D2}}$: (a) $v_{\mathrm{A1}}$ and $m_{\mathrm{s1}}$, for $v_{\mathrm{D}1}$=1 Hz/s; (b) $v_{\mathrm{A2}}$ and $m_{\mathrm{s2}}$, for $v_{\mathrm{D}2}$=1 $\mu$Hz/s$^{2}$.}
\label{fig:5}
\end{figure}

The relations among $v_{\mathrm{D0}}$, $v_{\mathrm{A0}}$, and $m_{\mathrm{s0}}$ have now been established. Because $v_{\mathrm{D1}}$ and $v_{\mathrm{D2}}$ are, respectively, the rate of change and acceleration of $v_{\mathrm{D0}}$, their responses can be derived from the $v_{\mathrm{D0}}$ result. The signals generated at point A are denoted by $v_{\mathrm{A1}}$ and $v_{\mathrm{A2}}$, and continuous integration gives the measurements, $m_{\mathrm{s1}}$ and $m_{\mathrm{s2}}$. Recovering the corresponding Doppler-frequency contributions requires one and two time integrations, respectively. Defining $r=t-c\tau$ for compact notation gives:
\begin{equation*}
m_{\mathrm{s1}}(t)=\int_0^t v_{\mathrm{A1}}(x)dx
=
\begin{cases}
\dfrac{v_{\mathrm{D1}}}{ga\tau}
\left[
\tau c(c+1)/2+r(c+1)
\right],
& \text{(initial stage)}
\\[8pt]
\dfrac{v_{\mathrm{D1}}}{2ga\tau}t^2,
& \text{(steady stage )}
\end{cases}
\tag{14}\label{eq:14}
\end{equation*}
\begin{equation*}
m_{\mathrm{s2}}(t)=\int_0^t v_{\mathrm{A2}}(x)dx
=
\begin{cases}
\dfrac{v_{\mathrm{D2}}}{2ga\tau}
\left[
\tau^2 c(c+1)(2c+1)/6
+\tau cr(c+1)
+r^2(c+1)
\right],
& \text{(initial stage)}
\\[8pt]
\dfrac{v_{\mathrm{D2}}}{6ga\tau}t^3,
& \text{(steady stage )}
\end{cases}
\tag{15}\label{eq:15}
\end{equation*}
The waveforms of $v_{\mathrm{A1}}$, $v_{\mathrm{A2}}$, $m_{\mathrm{s1}}$, and $m_{\mathrm{s2}}$ are plotted in Fig.~\ref{fig:5}.

Because the responses $v_{\mathrm{A0}}$, $v_{\mathrm{A1}}$, and $v_{\mathrm{A2}}$ coexist at point A and are transformed by the integrator into $m_{\mathrm{s0}}$, $m_{\mathrm{s1}}$, and $m_{\mathrm{s2}}$, the complete measurement is $m_{\mathrm{s}}=m_{\mathrm{s0}}+m_{\mathrm{s1}}+m_{\mathrm{s2}}$.

\subsubsection{Measurement of signals in the dual-arm-locking}

In a dual-arm-locking, the quantities to be measured include common- and differential-mode signals, as shown in Fig.~\ref{fig:6}. The blocks $P_{12}(s)$ and $P_{13}(s)$ represent the two arm sensors, and $v_{\mathrm{D12}}$ and $v_{\mathrm{D13}}$ are the corresponding Doppler-frequencies. Their sum is the common-mode signal, $v_{\mathrm{D}+}$, and their difference is the differential-mode signal, $v_{\mathrm{D}-}$. The two phasemeter outputs, $p_{12}$ and $p_{13}$, are fed to both an adder and a subtractor. The outputs, $v_{+}$ and $v_{-}$, therefore carry information about $v_{\mathrm{D}+}$ and $v_{\mathrm{D}-}$, respectively. Integrating $v_{+}$ and $v_{-}$ gives the measurement signals, $m_{+}$ and $m_{-}$. The roles of $v_{+}$, $v_{-}$ and $m_{+}$, $m_{-}$ are analogous to those of $v_{\mathrm{A}}$ and $m_{\mathrm{s}}$ in the single-arm-locking.

\begin{figure}[H]
\centering
\includegraphics[width=0.50\textwidth]{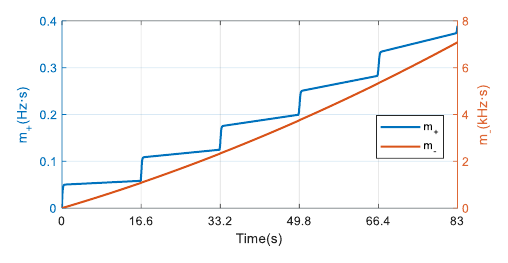}
\caption{Common- and differential-mode measurement signals, $m_{+}$ and $m_{-}$, for $g$=100, $a$=10, $\bar{\tau}$=16.6 s, $\Delta\tau$=62 ms, $v_{\mathrm{D+0}}$=100 Hz, $v_{\mathrm{D-0}}$=60 Hz, $v_{\mathrm{D+1}}$=1 Hz/s, $v_{\mathrm{D-1}}$=0.6 Hz/s, $v_{\mathrm{D+2}}$=1 $\mu$Hz/s$^{2}$, and $v_{\mathrm{D-2}}$=0.6 $\mu$Hz/s$^{2}$.}
\label{fig:7}
\end{figure}

In Fig.~\ref{fig:6}, the transfer function of Sensor 1$x$ ($x$=2 or 3) is denoted by $P_{1x}(s)$:

\begin{figure}[!htbp]
\centering
\includegraphics[width=0.50\textwidth]{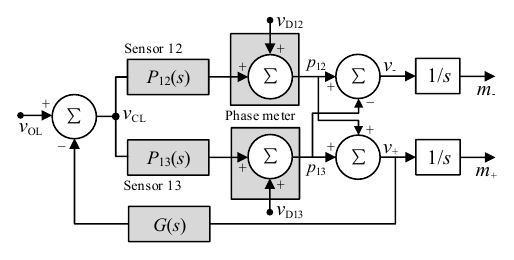}
\caption{Measurement model for the common- and differential-mode Doppler-frequency signals.}
\label{fig:6}
\end{figure}
\begin{equation*}
P_{1x}(s)=1-e^{-s\tau_{1x}}
\tag{16}\label{eq:16}
\end{equation*}
With $v_{\mathrm{D12}}$ and $v_{\mathrm{D13}}$ as model inputs, the output $v_{+}$ is:
\begin{equation*}
v_+(s)=\frac{v_{\mathrm{D12}}+v_{\mathrm{D13}}}{1+[P_{12}(s)+P_{13}(s)]G(s)}
 =\frac{v_{\mathrm{D+}}}{1+P_+(s)G(s)}
\tag{17}\label{eq:17}
\end{equation*}
where $P_{+}(s)$ is the common-mode sensor transfer function. Define $\bar{\tau}=(\tau_{12}+\tau_{13})/2$ and $\Delta\tau=(\tau_{12}-\tau_{13})/2$. Thus, $\bar{\tau}$ and $\Delta\tau$ are the mean delay and half-difference of $\tau_{12}$ and $\tau_{13}$, respectively, and the transfer function $P_{+}(s)$ is:
\begin{equation*}
P_+(s)=P_{12}(s)+P_{13}(s)=2-e^{-s\tau_{12}}-e^{-s\tau_{13}}=2\left[1-e^{-s\bar\tau}\cosh(s\Delta\tau)\right]
\tag{18}\label{eq:18}
\end{equation*}
where $\cosh(\cdot)$ is the hyperbolic cosine. Considering that the magnitude of $\Delta\tau$ is generally less than 0.5s, Eq.~\eqref{eq:18} can be further simplified as follows:
\begin{equation*}
P_+(s)\approx2\left(1-e^{-s\bar\tau}\right)
\tag{19}\label{eq:19}
\end{equation*}
The constant, 1st-order, and 2nd-order terms of the common-mode Doppler signal are denoted by $v_{\mathrm{D+0}}$, $v_{\mathrm{D+1}}$, and $v_{\mathrm{D+2}}$, with corresponding measurements $m_{+0}$, $m_{+1}$, and $m_{+2}$. Because $v_{+0}$ and $P_{+}(s)$ have the same functional forms as $v_{\mathrm{A0}}$ and $P_{\mathrm{s}}(s)$ in the single-arm-locking, the expressions for $m_{+}$ and $m_{\mathrm{s}}$ are analogous. For compact notation, define the general function $m_{\mathrm{c}}$ as:
\begin{equation*}
m_{\mathrm c}(t,v)=\begin{cases}
 \dfrac{v}{2ga}\left\lfloor\dfrac{t}{\bar\tau}\right\rfloor,&\text{(initial stage)}\\[6pt]
 \dfrac{v}{2ga\bar\tau}t,&\text{(steady stage )}
 \end{cases}
\tag{20}\label{eq:20}
\end{equation*}
The common-mode measurement is then:
\begin{equation*}
m_+(t)=m_{+0}+m_{+1}+m_{+2}=m_{\mathrm c}(t,v_{\mathrm{D+0}})+\int_0^t m_{\mathrm c}(x,v_{\mathrm{D+1}})\,dx+\int_0^t\int_0^{y}m_{\mathrm c}(t'',v_{\mathrm{D+2}})\,dxdy
\tag{21}\label{eq:21}
\end{equation*}
In Fig.~\ref{fig:6}, $v_{+}$ is used as the controller input and forms negative feedback that suppresses $v_{\mathrm{D}+}$. Although $v_{-}$ is not used directly for control, it also contains Doppler information and must be analyzed. To derive the measurement of $v_{\mathrm{D}-}$, the two phasemeter outputs are first written as:
\begin{equation*}
\begin{cases}p_{12}=\dfrac{v_{\mathrm{D12}}[1+P_{13}(s)G(s)]-v_{\mathrm{D13}}P_{12}(s)G(s)}{1+[P_{12}(s)+P_{13}(s)]G(s)}\\[8pt]p_{13}=\dfrac{v_{\mathrm{D13}}[1+P_{12}(s)G(s)]-v_{\mathrm{D12}}P_{13}(s)G(s)}{1+[P_{12}(s)+P_{13}(s)]G(s)}\end{cases}
\tag{22}\label{eq:22}
\end{equation*}
The differential signal is:
\begin{equation*}
v_-=p_{12}-p_{13}=\frac{v_{\mathrm{D12}}[1+2P_{13}(s)G(s)]-v_{\mathrm{D13}}[1+2P_{12}(s)G(s)]}{1+[P_{12}(s)+P_{13}(s)]G(s)}
\tag{23}\label{eq:23}
\end{equation*}
Because $G(s)\gg1$, Eq.~\eqref{eq:23} can be approximated as:
\begin{equation*}
v_-=2\frac{v_{\mathrm{D12}}P_{13}(s)-v_{\mathrm{D13}}P_{12}(s)}{P_{12}(s)+P_{13}(s)}
\tag{24}\label{eq:24}
\end{equation*}
In the low-frequency range, $P_{12}(s)$ and $P_{13}(s)$ can be approximated by $s\tau_{12}$ and $s\tau_{13}$, respectively. Therefore, the simplified form of Eq.~\eqref{eq:24} can be further reduced to:
\begin{equation*}
v_- =
2\frac{v_\mathrm{D12}s\tau_{13}-v_\mathrm{D13}s\tau_{12}}
{s\tau_{12}+s\tau_{13}}
=
v_\mathrm{D12}-v_\mathrm{D13}
-\frac{\Delta\tau}{\bar{\tau}}
(v_\mathrm{D12}+v_\mathrm{D13})
=
v_\mathrm{D-}
-\frac{\Delta\tau}{\bar{\tau}}v_\mathrm{D+}
\tag{25}\label{eq:25}
\end{equation*}
Because $|\Delta\tau|\ll\bar{\tau}$, Eq.~\eqref{eq:25} can be further simplified as:
\begin{equation*}
v_-\approx v_{\mathrm{D-}}
\tag{26}\label{eq:26}
\end{equation*}
For consistency with the calculation of $m_{+}$, $v_{-}$ is also integrated, yielding:
\begin{equation*}
m_-(t)=\int_0^t v_-\,dx=t\,v_{\mathrm{D-}}
\tag{27}\label{eq:27}
\end{equation*}
The resulting $m_{+}$ and $m_{-}$ curves are shown in Fig.~\ref{fig:7}.

\subsection{Doppler-frequency estimation model}

\begin{figure}[H]
\centering
\includegraphics[width=0.67\textwidth]{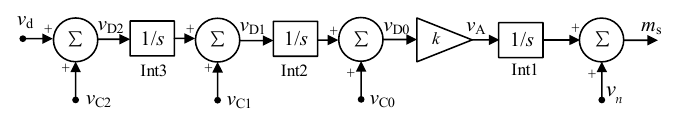}
\caption{Doppler-frequency estimation model.}
\label{fig:8}
\end{figure}

The preceding analysis establishes a Doppler-frequency measurement model for arm-locking. Laser noise and variations in spacecraft gravitational acceleration nevertheless introduce measurement errors, making accurate direct measurement difficult. To improve the estimate with a KF, we construct the model shown in Fig.~\ref{fig:8} using the single-arm-locking as an example. It contains several adders, integrators Int1-Int3, and an attenuator. The adders and Int2-Int3 represent dynamics implicit in the spacecraft motion rather than physical components, whereas Int1 is a deliberately introduced physical integrator corresponding to the integrator in Fig.~\ref{fig:2}. Because the Doppler-frequency is attenuated by a factor $1/ga\tau$ when measured at point A, the arm-locking is modeled as an attenuator with coefficient $k$=1/$ga\tau$. The initial values of $v_{\mathrm{D0}}$-$v_{\mathrm{D2}}$ are denoted $v_{\mathrm{C0}}$-$v_{\mathrm{C2}}$ and used as model inputs. Time derivatives above second order are represented by a disturbance $v_{\mathrm{d}}$. The output is $m_{\mathrm{s}}$ with additive measurement noise $v_{\mathrm{n}}$. The purpose of the model is to infer the instantaneous $v_{\mathrm{D0}}$-$v_{\mathrm{D2}}$ from the noisy $m_{\mathrm{s}}$ measurement.

Because the estimation model has multiple inputs, it is written in state-space form. The values of Int1, Int2, and Int3 are selected as the states: $x_{1}$, $x_{2}$, and $x_{3}$. Although $v_{\mathrm{D2}}$ is not the state of an inertial element, it varies slowly and is therefore introduced as the augmented state, $x_{4}$. $x_{1}$-$x_{4}$ form the state vector: $\bm{x}=$[$x_{1}$ $x_{2}$ $x_{3}$ $x_{4}$]$^{\mathrm{T}}$. $v_{\mathrm{C0}}$, $v_{\mathrm{C1}}$, and $v_{\mathrm{C2}}$ form the input vector: $\bm{u}$=[$u_{1}$ $u_{2}$ $u_{3}$]$^{\mathrm{T}}$ =[$v_{\mathrm{C0}}$ $v_{\mathrm{C1}}$ $v_{\mathrm{C2}}$]$^{\mathrm{T}}$. The observed variable is $\bm{y}=x_{1}=m_{\mathrm{s}}$. The differential equations obtained from the measurement model are:
\begin{equation*}
\dot{x}_1=kx_2+ku_1,\qquad \dot{x}_2=x_3+u_2,\qquad \dot{x}_3=x_4+u_3,\qquad \dot{x}_4=v_{\mathrm d}
\tag{28}\label{eq:28}
\end{equation*}
The measurement model can therefore be written in state-space form as:
\begin{equation*}
\begin{cases}
 \dot{\bm{x}}=\bm{A}_{\mathrm c}\bm{x}+\bm{B}_{\mathrm c}\bm{u}+\bm{w}\\
 \bm{y}=\bm{H}_{\mathrm c}\bm{x}+\bm{v}
\end{cases}
\tag{29}\label{eq:29}
\end{equation*}
where $\bm{A}_{\mathrm{c}}$, $\bm{B}_{\mathrm{c}}$, and $\bm{H}_{\mathrm{c}}$ are the state, input, and observation matrices, respectively:
\begin{equation*}
\bm{A}_{\mathrm c}=\begin{bmatrix}0&k&0&0\\0&0&1&0\\0&0&0&1\\0&0&0&0\end{bmatrix},\qquad
 \bm{B}_{\mathrm c}=\begin{bmatrix}k&0&0\\0&1&0\\0&0&1\\0&0&0\end{bmatrix},\qquad
 \bm{H}_{\mathrm c}=\begin{bmatrix}1&0&0&0\end{bmatrix}
\tag{30}\label{eq:30}
\end{equation*}
Here $\bm{w}$ is process noise with covariance $\bm{Q}_{\mathrm C}=\operatorname{diag}(0,0,0,q_{\mathrm p})$, where $q_{\mathrm{p}}$ is the variance of $v_{\mathrm{d}}$. The measurement noise $v_{\mathrm{n}}$ has covariance $\bm{R}$. Because the observation is scalar, $\bm{R}$ equals the variance of the measurement noise. The quantities $v_{\mathrm{D0}}$-$v_{\mathrm{D2}}$ to be estimated are obtained from:
\begin{equation*}
v_{\mathrm{D0}}=x_2+u_1=\dot{m}_{\mathrm s}/k,\qquad v_{\mathrm{D1}}=x_3+u_2=\ddot{m}_{\mathrm s}/k,\qquad v_{\mathrm{D2}}=x_4+u_3=\dddot{m}_{\mathrm s}/k
\tag{31}\label{eq:31}
\end{equation*}
Eq.~\eqref{eq:31} shows that the individual Doppler-frequency components can be inferred using only $m_{\mathrm{s}}$.

The same estimation model applies to dual-arm-locking. For the common-mode signal, Eq.~\eqref{eq:20} gives $k=1/(2ga\bar{\tau})$; for the differential-mode signal, Eq.~\eqref{eq:26} gives $k$=1.

\subsection{Doppler-frequency suppression model}

\begin{figure}[!htbp]
\centering
\includegraphics[width=0.50\textwidth]{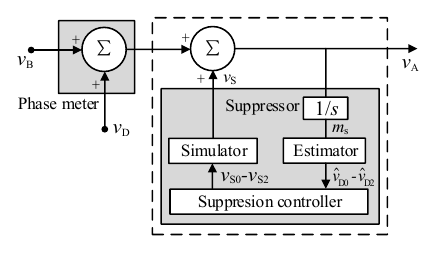}
\caption{Structure of the Doppler-frequency suppressor.}
\label{fig:9}
\end{figure}

To real-time suppress the Doppler-frequency-pulling, the Doppler-frequency suppressor shown in Fig.~\ref{fig:9} is introduced. Taking the single-arm-locking in Fig.~\ref{fig:2} as an example, the suppressor is inserted at the phasemeter output to compensate the Doppler signal entering the loop. The estimator uses $v_{\mathrm{A}}$ to estimate the constant component, $\hat v_{\mathrm{D0}}$, the 1st-order time derivative, $\hat v_{\mathrm{D1}}$, and the 2nd-order component, $\hat v_{\mathrm{D2}}$ of $v_{\mathrm{D}}$. The controller then takes $\hat v_{\mathrm{D0}}$-$\hat v_{\mathrm{D2}}$ as its input and generates $v_{\mathrm{S0}}$-$v_{\mathrm{S2}}$ to suppress $v_{\mathrm{D0}}$-$v_{\mathrm{D2}}$, respectively. These components are combined by the Doppler-frequency simulator to form the complete suppression signal, $v_{\mathrm{S}}$, which is injected into the loop to cancel $v_{\mathrm{D}}$ and thereby suppress frequency-pulling.

\begin{figure}[!htbp]
\centering
\includegraphics[width=0.50\textwidth]{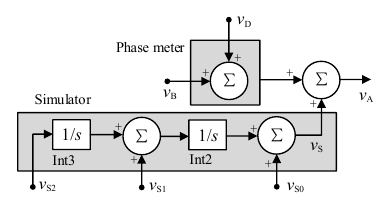}
\caption{Structure of the Doppler-frequency simulator.}
\label{fig:10}
\end{figure}

The Doppler-frequency simulator generates a synthetic Doppler signal, as shown in Fig.~\ref{fig:10}. It consists of integrators, Int2 and Int3, and two adders. The inputs $v_{\mathrm{S0}}$-$v_{\mathrm{S2}}$ are combined to produce the  $v_{\mathrm{S}}$, which is added downstream of the phasemeter to compensate $v_{\mathrm{D}}$.

\begin{figure}[H]
\centering
\includegraphics[width=0.62\textwidth]{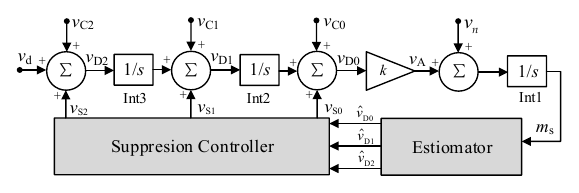}
\caption{Complete Doppler-frequency suppressor model.}
\label{fig:11}
\end{figure}

Combining the estimation model in Fig.~\ref{fig:8}, the suppressor structure in Fig.~\ref{fig:9}, and the simulator in Fig.~\ref{fig:10} yields the complete suppressor model in Fig.~\ref{fig:11}. The Int2 and Int3 elements in the estimation model are merged with their counterparts in the simulator. The suppressor inputs are $v_{\mathrm{C0}}$-$v_{\mathrm{C2}}$, and its purpose is to prevent growth of $v_{\mathrm{A}}$ and drive it toward zero.

Comparison of the estimation and suppression models, excluding the estimator and controller, shows that they are identical except for the internal control signals $v_{\mathrm{S0}}$-$v_{\mathrm{S2}}$. The suppression-model input vector therefore becomes $\bm{u}$=[$v_{\mathrm{S0}}$, $v_{\mathrm{S1}}$, $v_{\mathrm{S2}}$]$^{\mathrm{T}}$, whereas the state vector and the state, input, and observation matrices remain unchanged.

\section{Method}

This section presents the arm-locking controller and suppressor designs. We uses a PI controller combined with a high-pass filter as the Arm-locking controller, while the suppressor contains a state estimator and a suppression controller.

\subsection{Arm-locking controller}

\subsubsection{Controller design}

The PI arm-locking controller introduced in Ref. \cite{zhang2024} is simple and provides sufficient noise suppression, but its behavior with respect to Doppler-frequency-pulling was not analyzed there. Section II showed in detail how pulling arises when a PI controller is used for arm-locking. Fig.~\ref{fig:2} shows that the frequency-pulling continues to grow linearly with time in the Initial-state, its expression is:
\begin{equation*}
v_{\mathrm P}(s)=\frac{-v_{\mathrm D}G(s)}{1+P(s)G(s)}
 =\frac{-v_{\mathrm D}g(s+a)}{s+g(s+a)(1-e^{-s\tau})}
\tag{32}\label{eq:32}
\end{equation*}
As $t$ approaches infinity ($s$ approaches zero), the magnitude of $v_{\mathrm{P}}$ diverges, showing that the PI arm-locking controller alone cannot suppress frequency-pulling. To provide frequency-pulling suppression, a 2nd-order high-pass filter with transfer function $F_{\mathrm{H}}(s)$ is added to the loop:
\begin{equation*}
F_{\mathrm H}(s)=\frac{s^2}{(s+p_{\mathrm a})(s+p_{\mathrm b})}
\tag{33}\label{eq:33}
\end{equation*}
where $p_{\mathrm{a}}$ and $p_{\mathrm{b}}$ are the filter poles. The arm-locking controller formed by the PI controller and the filter has transfer function:
\begin{equation*}
G(s)=g\frac{s(s+a)}{(s+p_{\mathrm a})(s+p_{\mathrm b})}
\tag{34}\label{eq:34}
\end{equation*}
The corresponding frequency pulling is:
\begin{equation*}
v_{\mathrm P}(s)=\frac{-v_{\mathrm D}g(s+a)s}{(s+p_{\mathrm a})(s+p_{\mathrm b})+g(s+a)s(1-e^{-s\tau})}
\tag{35}\label{eq:35}
\end{equation*}
As $s$ approaches zero, $v_{\mathrm{P}}$ approaches zero, therefore the 2nd-order high-pass filter removes the DC pulling contribution.

\subsubsection{Stability analysis}

Adding the filter changes the open-loop phase and gain, therefore requires a stability analysis. We consider both time- and frequency-domain criteria. The time-domain analysis follows Ref. \cite{zhang2024}. The modified arm-locking controller, $G'(s)$, obtained by cascading the PI controller, $G(s)$, with the high-pass filter, $F_{\mathrm{H}}(s)$, is:
\begin{equation*}
G'(s)=g\frac{B(s)}{A(s)}=g\frac{s(s+a)}{(s+p_{\mathrm a})(s+p_{\mathrm b})}
\tag{36}\label{eq:36}
\end{equation*}
Define the function required by the stability criterion as:
\begin{equation*}
H_{G'}(s)=\frac{gB(s)}{A(s)+gB(s)}=\frac{gs(s+a)}{(s+p_{\mathrm a})(s+p_{\mathrm b})+gs(s+a)}
\tag{37}\label{eq:37}
\end{equation*}
The system is stable if the following two conditions are satisfied:\\
(1) All poles of $H_{\mathrm{G'}}(s)$ have negative real parts:
\begin{equation*}
\operatorname{Re}(p_i)<0,\quad i=1,2,\ldots,r
\tag{38A}\label{eq:38A}
\end{equation*}
(2) The magnitude response of $H_{\mathrm{G'}}(s)$ does not exceed unity:
\begin{equation*}
\max_{\omega\in[-\infty,+\infty]}\left|H_{G'}(j\omega)\right|\le1
\tag{38B}\label{eq:38B}
\end{equation*}
$H_{\mathrm{G'}}(s)$ has two poles, denoted as $p_{1}$ and $p_{2}$:
\begin{equation*}
p_{1,2}=\frac{-B\pm\sqrt{B^2-4(1+g)p_{\mathrm a}p_{\mathrm b}}}{2(1+g)},\qquad B=p_{\mathrm a}+p_{\mathrm b}+ga
\tag{39}\label{eq:39}
\end{equation*}
Because $a>0$, $g>0$, $p_{\mathrm{a}}>0$, and $p_{\mathrm{b}}>0$, the real parts of both $p_{1}$ and $p_{2}$ are negative, satisfying condition (38A).

To test the second condition, first expand $A(s)$ in partial fractions:
\begin{equation*}
A(s)=C_0+\frac{C_1}{s-p_1}+\frac{C_2}{s-p_2},\quad C_0=\frac{1}{1+g},\quad C_1=\frac{(p_{\mathrm b}-p_{\mathrm a}-a)p_1-ap_{\mathrm a}}{(1+g)^2(p_1-p_2)},\quad C_2=\frac{(p_{\mathrm b}-p_{\mathrm a}-a)p_2-ap_{\mathrm a}}{(1+g)^2(p_2-p_1)}
\tag{40}\label{eq:40}
\end{equation*}
The quantity required by the criterion is then:
\begin{equation*}
D=C_0-\frac{C_1}{p_1}-\frac{C_2}{p_2}=0
\tag{41}\label{eq:41}
\end{equation*}
When $D\le1$, the maximum magnitude of $H_{\mathrm{G'}}(s)$ is below unity, so condition (38B) is satisfied and the system is stable.

\begin{figure}[!htbp]
\centering
\begin{minipage}[t]{0.49\textwidth}
\centering
\includegraphics[width=\linewidth]{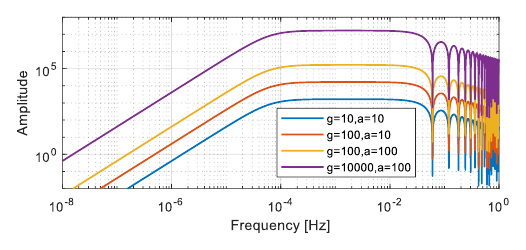}
\par\smallskip (a)
\end{minipage}
\hfill
\begin{minipage}[t]{0.49\textwidth}
\centering
\includegraphics[width=\linewidth]{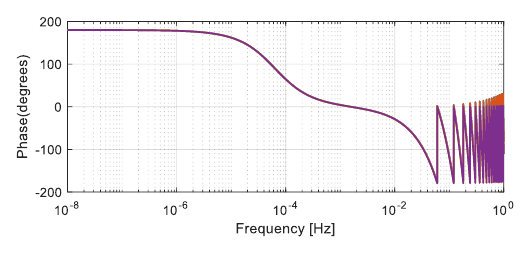}
\par\smallskip (b)
\end{minipage}
\caption{Open-loop Bode plot of the single-arm-locking for different combinations of $g$ and $a$, $\tau$=16.6 s. (a) Amplitude-frequency response curves; (b) Phase-frequency response curves.}
\label{fig:12}
\end{figure}

The frequency-domain stability is assessed from the open-loop Bode plot in Fig.~\ref{fig:12}. The amplitude response is approximately flat from 0.1 mHz to 60 mHz, and a larger product $g\times a$ produces a larger open-loop gain. At sufficiently low frequency the gain decreases and eventually crosses unity. The phase remains within -180 deg to +180 deg over the full range. Below the first sensor null, the phase curves nearly overlap, showing that the phase response is almost independent of $g$ and $a$. At each unity-gain crossover, the phase excursion does not exceed 180 deg, consistent with stable operation.

\subsubsection{Noise-suppression capability}

\begin{figure}[!htbp]
\centering
\includegraphics[width=0.535\textwidth]{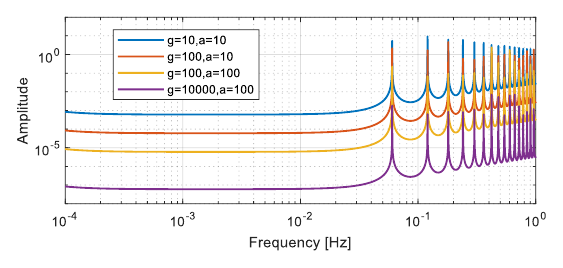}
\caption{Frequency-noise suppression curves for different combinations of $g$ and $a$.}
\label{fig:13}
\end{figure}

Laser-frequency-noise suppression is a key arm-locking performance metric. In Fig.~\ref{fig:2}, $v_{\mathrm{OL}}$ is the input and $v_{\mathrm{CL}}$ is the output; the corresponding transfer function is:
\begin{equation*}
H_{\mathrm{CL}}=\frac{1}{1+G'(s)(1-e^{-s\tau})}
\tag{42}\label{eq:42}
\end{equation*}
To evaluate the suppression response, specific controller parameters are selected. For the high-pass filter, set $p_{\mathrm{a}}=p_{\mathrm{b}}$= 4$\times$10$^{-4}$ rad/s, corresponding to a cutoff near the lower edge of the science band (0.1 mHz). For the PI controller, the parameters $g$ and $a$ jointly determine the suppression capability. The amplitude response is shown in Fig.~\ref{fig:13}.

From 0.1 mHz to the first arm-sensor null at approximately 60 mHz, the suppression response is nearly flat. Increasing $g$ and $a$ will increases the suppression strength.

\subsection{Suppressor design}

The suppressor is designed first to real-time estimate $v_{\mathrm{D}}$ and then to suppress it by feedback. Because estimation is performed in the presence of strong laser noise, a Kalman filter (KF), which provides optimal linear estimation in a stochastic environment, is adopted. Immediately after arm-locking loop closure, $v_{\mathrm{D}}$ can produce large frequency pulling, so an LQR controller is used to drive the pulling rapidly toward zero. The combination of the KF and LQR forms an LQG controller; accordingly, the proposed suppressor uses LQG controller for simultaneous real-time estimation and suppression of $v_{\mathrm{D}}$.

\begin{figure}[H]
\centering
\includegraphics[width=0.50\textwidth]{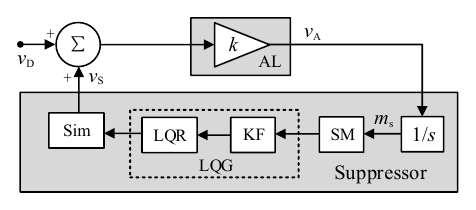}
\caption{Internal structure of the suppressor.}
\label{fig:14}
\end{figure}

For the single-arm-locking, Fig.~\ref{fig:14} shows the connection between the suppressor and the arm-locking (AL) as well as the internal suppressor structure. The arm-locking is represented by an attenuator with coefficient $k$. The suppressor consists of an integrator $(1/s)$, an arithmetic-mean smoother (SM), the LQG controller, and a Doppler-frequency simulator (Sim).

After loop closure, the output of ALS($v_{\mathrm{A}}$) is integrated to obtain $m_{\mathrm{s}}$. To reduce the influence of high-frequency laser-noise components on the estimate, SM averages $m_{\mathrm{s}}$ over intervals of duration $\tau$. The smoothed measurement is supplied to the KF. The LQR generates a control signal from the current estimated state and sends it to Sim, which constructs a suppression signal($v_{\mathrm{S}}$) and injects it into the loop to suppress $v_{\mathrm{A}}$. When $v_{\mathrm{A}}$ approaches zero, $v_{\mathrm{P}}$ also approaches zero, thereby suppressing frequency pulling.

\subsubsection{Discretization of the measurement model}

For optimal estimation of $v_{\mathrm{D}}$ in a strong-noise environment, a filtering model must be constructed before applying the KF. The filtering model is obtained by discretizing the continuous measurement model in Eq.~\eqref{eq:29}. The arm round-trip time, $\tau$, is used as the KF sampling interval $T_{\mathrm{s}}$, and a zero-order hold is used for discretization, giving:
\begin{equation*}
\left\{
\begin{aligned}
\bm{x}_k &= \bm{A}_d\bm{x}_{k-1}
+\bm{B}_d\bm{u}_{k-1}
+\bm{w}_k,\\
\bm{y}_k &= \bm{H}_d\bm{x}_k+\bm{v}_k
\end{aligned}
\right.
\tag{43}\label{eq:43}
\end{equation*}
where $\bm{A}_{\mathrm{d}}$, $\bm{B}_{\mathrm{d}}$, and $\bm{H}_{\mathrm{d}}$ are the discrete state-transition, input, and observation matrices, obtained from:
\begin{equation*}
\bm{A}_{\mathrm d}=e^{\bm{A}_{\mathrm c}T_{\mathrm s}},\qquad
 \bm{B}_{\mathrm d}=\int_0^{T_{\mathrm s}}e^{\bm{A}_{\mathrm c}t}\bm{B}_{\mathrm c}\,dt,\qquad
 \bm{H}_{\mathrm d}=\bm{H}_{\mathrm c}
\tag{44}\label{eq:44}
\end{equation*}
The corresponding matrix elements are:
\begin{equation*}
\bm{A}_{\mathrm d}=\begin{bmatrix}1&kT_{\mathrm s}&kT_{\mathrm s}^2/2&kT_{\mathrm s}^3/6\\0&1&T_{\mathrm s}&T_{\mathrm s}^2/2\\0&0&1&T_{\mathrm s}\\0&0&0&1\end{bmatrix},\quad\bm{B}_{\mathrm d}=\begin{bmatrix}kT_{\mathrm s}&kT_{\mathrm s}^2/2&kT_{\mathrm s}^3/6\\0&T_{\mathrm s}&T_{\mathrm s}^2/2\\0&0&T_{\mathrm s}\\0&0&0\end{bmatrix},\quad\bm{H}_{\mathrm d}=\begin{bmatrix}1&0&0&0\end{bmatrix}
\end{equation*}
The discrete process-noise covariance is denoted $\bm{Q}_{\mathrm{d}}$ and is given by:
\begin{equation*}
\bm{Q}_{\mathrm d}=\int_0^{T}\exp(\bm{A}_{\mathrm c}t)\bm{Q}_{\mathrm c}\exp(\bm{A}_{\mathrm c}^{\mathrm T}t)\,dt
\tag{45}\label{eq:45}
\end{equation*}

\subsubsection{Kalman filter}

In the state space, the Kalman filter is applied recursively. Suppose that at time $k$-$1$ the filtered estimate, $\hat{\bm{x}}_{k-1}$, and its error covariance, $\bm{P}_{k-1}$, are available. The prediction step is:
\begin{equation*}
\begin{cases}\hat{\bm{x}}_{k|k-1}=\bm{A}_{\mathrm d}\hat{\bm{x}}_{k-1}+\bm{B}_{\mathrm d}\bm{u}_{k-1}\\\bm{P}_{k|k-1}=\bm{A}_{\mathrm d}\bm{P}_{k-1}\bm{A}_{\mathrm d}^{\mathrm T}+\bm{Q}_{\mathrm d}\end{cases}
\tag{46}\label{eq:46}
\end{equation*}
The Kalman gain is:
\begin{equation*}
\bm{K}_{\mathrm d,k}=\bm{P}_{k|k-1}\bm{H}_{\mathrm d}^{\mathrm T}\left(\bm{H}_{\mathrm d}\bm{P}_{k|k-1}\bm{H}_{\mathrm d}^{\mathrm T}+\bm{R}\right)^{-1}
\tag{47}\label{eq:47}
\end{equation*}
Using the current observation, $\bm{y}_{k}$, the innovation is:
\begin{equation*}
\bm{e}_k=\bm{y}_k-\bm{H}_{\mathrm d}\hat{\bm{x}}_{k|k-1}
\tag{48}\label{eq:48}
\end{equation*}
and the corrected state estimate is:
\begin{equation*}
\hat{\bm{x}}_k=\hat{\bm{x}}_{k|k-1}+\bm{K}_{\mathrm d,k}\bm{e}_k
\tag{49}\label{eq:49}
\end{equation*}
The covariance is updated according to:
\begin{equation*}
\bm{P}_k=(\bm{I}-\bm{K}_{\mathrm d,k}\bm{H}_{\mathrm d})\bm{P}_{k|k-1}(\bm{I}-\bm{K}_{\mathrm d,k}\bm{H}_{\mathrm d})^{\mathrm T}+\bm{K}_{\mathrm d,k}\bm{R}\bm{K}_{\mathrm d,k}^{\mathrm T}
\tag{50}\label{eq:50}
\end{equation*}

\subsubsection{LQR controller}

After the estimator provides the state estimate, a LQR is used for control. LQR is a full-state feedback method in which the feedback coefficients are obtained by solving a Riccati equation. In the suppressor model, $x_{4}$ is a slowly varying but uncontrollable state. Including $x_{4}$ directly makes the Riccati equation unsolvable. The subsystem formed by $x_{1}$-$x_{3}$ is fully controllable, however, and the estimate of $x_{4}$ can be sign-reversed and added to $u_{3}$ as a feedforward term to cancel its long-term effect on the integration chain. The suppression controller therefore uses a combined feedforward-plus-feedback law.

The controllable subsystem is regulated by the LQR and obeys:
\begin{equation*}
\bm{x}_{\mathrm L,k+1}=\bm{A}_{\mathrm L}\bm{x}_{\mathrm L,k}+\bm{B}_{\mathrm L}\bm{u}_{\mathrm L,k}
\tag{51}\label{eq:51}
\end{equation*}
where $\bm{x}_{\mathrm{L}}$, $\bm{A}_{\mathrm{L}}$, and $\bm{B}_{\mathrm{L}}$ are the subsystem state vector, state-transition matrix, and input matrix, respectively:
\begin{equation*}
\bm{x}_{\mathrm L}=\begin{bmatrix}x_1\\x_2\\x_3\end{bmatrix},\qquad
 \bm{A}_{\mathrm L}=\begin{bmatrix}1&kT_{\mathrm s}&kT_{\mathrm s}^2/2\\0&1&T_{\mathrm s}\\0&0&1\end{bmatrix},\qquad
 \bm{B}_{\mathrm L}=\begin{bmatrix}kT_{\mathrm s}&kT_{\mathrm s}^2/2&kT_{\mathrm s}^3/6\\0&T_{\mathrm s}&T_{\mathrm s}^2/2\\0&0&T_{\mathrm s}\end{bmatrix}
\tag{52}\label{eq:52}
\end{equation*}
The LQR minimizes the cost function:
\begin{equation*}
J=\sum_{k=0}^{\infty}\left(\bm{x}_{\mathrm L,k}^{\mathrm T}\bm{Q}_{\mathrm L}\bm{x}_{\mathrm L,k}+\bm{u}_{\mathrm L,k}^{\mathrm T}\bm{R}_{\mathrm L}\bm{u}_{\mathrm L,k}\right)
\tag{53}\label{eq:53}
\end{equation*}
where $\bm{Q}_{\mathrm{L}}$ is the state-weighting matrix and $\bm{R}_{\mathrm{L}}$ is the control-weighting matrix. To drive the output toward zero, $x_{1}$ is assigned a comparatively large weight. Because $x_{2}$ and $x_{3}$ affect $x_{1}$ through the integration chain, they are also weighted appropriately. We choose:
\begin{equation*}
\bm{Q}_{\mathrm L}=\operatorname{diag}(q_1,q_2,q_3),\qquad q_1>q_2>q_3
\tag{54}\label{eq:54}
\end{equation*}
To avoid an excessively large $u_{3}$ and the resulting slow convergence, $u_{3}$ is assigned a larger control penalty, and we choose:
\begin{equation*}
\bm{R}_{\mathrm L}=\operatorname{diag}(r_1,r_2,r_3),\qquad r_3>r_1,\quad r_1\approx r_2
\tag{55}\label{eq:55}
\end{equation*}
After $\bm{Q}_{\mathrm{L}}$ and $\bm{R}_{\mathrm{L}}$ are specified, the discrete Riccati equation is solved:
\begin{equation*}
\bm{P}_{\mathrm L}=\bm{Q}_{\mathrm L}+\bm{A}_{\mathrm L}^{\mathrm T}\bm{P}_{\mathrm L}\bm{A}_{\mathrm L}-\bm{A}_{\mathrm L}^{\mathrm T}\bm{P}_{\mathrm L}\bm{B}_{\mathrm L}\left(\bm{R}_{\mathrm L}+\bm{B}_{\mathrm L}^{\mathrm T}\bm{P}_{\mathrm L}\bm{B}_{\mathrm L}\right)^{-1}\bm{B}_{\mathrm L}^{\mathrm T}\bm{P}_{\mathrm L}\bm{A}_{\mathrm L}
\tag{56}\label{eq:56}
\end{equation*}
Substituting the solution $\bm{P}_{\mathrm{L}}$ gives the optimal feedback gain:
\begin{equation*}
\bm{K}_{\mathrm L}=\left(\bm{R}_{\mathrm L}+\bm{B}_{\mathrm L}^{\mathrm T}\bm{P}_{\mathrm L}\bm{B}_{\mathrm L}\right)^{-1}
 \bm{B}_{\mathrm L}^{\mathrm T}\bm{P}_{\mathrm L}\bm{A}_{\mathrm L}
\tag{57}\label{eq:57}
\end{equation*}
The output of LQR controller is:
\begin{equation*}
\bm{u}_{\mathrm L}=-\bm{K}_{\mathrm L}\hat{\bm{x}}_{\mathrm L}
\tag{58}\label{eq:58}
\end{equation*}
where, $\hat{x}_{\mathrm L}$ is the estimation of $\bm{x}_{\mathrm L}$. Combining this feedback with feedforward compensation of the estimated $x_{4}$ through $u_{3}$ gives the complete gain matrix $\bm{K}_{4}$ and hence:
\begin{equation*}
\bm{K}_{\mathrm f}=\begin{bmatrix}\bm{K}_{\mathrm L}&\bm{K}_4\end{bmatrix},\qquad
 \bm{K}_4=\begin{bmatrix}0&0&1\end{bmatrix}^{\mathrm T}
\tag{59}\label{eq:59}
\end{equation*}
The output of suppression controller is:
\begin{equation*}
\bm{u}=-\bm{K}_{\mathrm f}\hat{\bm{x}}
\tag{60}\label{eq:60}
\end{equation*}

\subsubsection{Stability analysis}

The LQR uses state estimates for closed-loop control. It cannot directly access the true state($\bm{x}$) and instead uses the state($\hat{\bm{x}}$) estimated by the KF. Because $\hat{\bm{x}}$ differs from $\bm{x}$ by the error, $\bm{e}$, the stability of the resulting closed-loop system must be examined. In the closed-loop system, although $\bm{x}_{4}$ is an uncontrollable augmented state, because it varies slowly and has been canceled by the feedforward control, only the feedback system consisting of the LQR and the KF needs to be considered for stability analysis. The control law is rewritten as:
\begin{equation*}
\bm{u}_{\mathrm L}=-\bm{K}_{\mathrm L}(\bm{x}_k-\bm{e}_k)
\tag{61}\label{eq:61}
\end{equation*}
Substituting this expression into the system equation and neglecting noise gives:
\begin{equation*}
\bm{x}_{k+1}=\bm{A}_{\mathrm d}\bm{x}_k-\bm{B}_{\mathrm d}\bm{K}_{\mathrm L}(\bm{x}_k-\bm{e}_k)
\tag{62}\label{eq:62}
\end{equation*}
The estimation error evolves according to:
\begin{equation*}
\bm{e}_{k+1}=(\bm{I}-\bm{K}_{\mathrm d}\bm{H}_{\mathrm d})\bm{A}_{\mathrm d}\bm{e}_k
\tag{63}\label{eq:63}
\end{equation*}
The closed-loop system containing $\bm{x}$ and $\bm{e}$ is therefore:
\begin{equation*}
\begin{bmatrix}\bm{x}_{k+1}\\\bm{e}_{k+1}\end{bmatrix}=\begin{bmatrix}\bm{A}_{\mathrm d}-\bm{B}_{\mathrm d}\bm{K}_{\mathrm L}&\bm{B}_{\mathrm d}\bm{K}_{\mathrm L}\\\bm{0}&(\bm{I}-\bm{K}_{\mathrm d}\bm{H}_{\mathrm d})\bm{A}_{\mathrm d}\end{bmatrix}\begin{bmatrix}\bm{x}_k\\\bm{e}_k\end{bmatrix}
\tag{64}\label{eq:64}
\end{equation*}
The closed-loop matrix is block upper triangular, so its eigenvalues, $\lambda_{\mathrm{LQG}}$ are the union of two sets:
\begin{equation*}
\lambda_{\mathrm{LQG}}=\lambda(\bm{A}_{\mathrm d}-\bm{B}_{\mathrm d}\bm{K}_{\mathrm L})\cup\lambda\!\left((\bm{I}-\bm{K}_{\mathrm d}\bm{H}_{\mathrm d})\bm{A}_{\mathrm d}\right)
\tag{65}\label{eq:65}
\end{equation*}
This is the LQG separation principle: the closed-loop poles consist of the LQR control poles and the KF estimation-error poles. According to the separation principle,\\
(1) If ($\bm{A}_{\mathrm{d}}$, $\bm{B}_{\mathrm{d}}$) is controllable and is observable, a LQR gain, $\bm{K}_{\mathrm{L}}$, exists such that all eigenvalues of $\bm{A}_{\mathrm{d}}-\bm{B}_{\mathrm{d}}\bm{K}_{\mathrm{L}}$ lie inside the unit circle, so the controlled dynamics are stable.\\
(2) At steady state, the Kalman gain approaches a constant, $\bm{K}_{\mathrm{\infty}}$, and all eigenvalues of $\bm{A}_{\mathrm{d}}-\bm{K}_{\mathrm{\infty}}\bm{H}_{\mathrm{d}}\bm{A}_{\mathrm{d}}$ also lie inside the unit circle, so the filtering-error dynamics are stable.

To apply the separation principle, controllability and observability are verified separately. The controllability matrix is:
\begin{equation*}
\bm{M}=\begin{bmatrix}\bm{B}_{\mathrm d}&\bm{A}_{\mathrm d}\bm{B}_{\mathrm d}&\bm{A}_{\mathrm d}^{2}\bm{B}_{\mathrm d}\end{bmatrix}
\tag{66}\label{eq:66}
\end{equation*}
Its calculated rank is rank($\bm{M}$)=3, equal to the state dimension, so the system is controllable. The observability matrix is:
\begin{equation*}
\bm{N}=\begin{bmatrix}
 \bm{H}_{\mathrm d}^{\mathrm T}&(\bm{H}_{\mathrm d}\bm{A}_{\mathrm d})^{\mathrm T}&(\bm{H}_{\mathrm d}\bm{A}_{\mathrm d}^2)^{\mathrm T}&(\bm{H}_{\mathrm d}\bm{A}_{\mathrm d}^3)^{\mathrm T}
 \end{bmatrix}^{\mathrm T}
\tag{67}\label{eq:67}
\end{equation*}
Its calculated rank is rank($\bm{N}$)=4, again equal to the augmented state dimension. Therefore, the system satisfies the separation principle, all poles of the augmented closed-loop system lie inside the open unit disk, and the system is asymptotically stable.

\subsubsection{Time-varying suppression strategy}

\begin{figure}[H]
\centering
\includegraphics[width=0.51\textwidth]{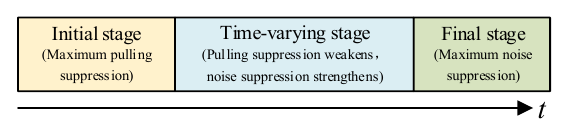}
\caption{Three stages of the time-varying suppression process.}
\label{fig:15}
\end{figure}

After loop closure, if Doppler-frequency-pulling is not suppressed, the pulling grows during each light-propagation cycle, so rapid mitigation is essential. Strong pulling suppression, however, substantially weakens laser-frequency-noise suppression. Therefore, once pulling has been brought under control, the pulling-suppression strength should be reduced gradually while the noise-suppression capability is increased until the TDI requirement is reached. The strategy in Fig.~\ref{fig:15} divides the operation into an initial stage, a time-varying stage, and a final stage. Pulling suppression is strongest initially; during the time-varying-stage it is gradually reduced while noise suppression is strengthened; and in the final stage the system provides its strongest noise suppression.

The suppressor strength depends on $\bm{Q}_{\mathrm{L}}$ and $\bm{R}_{\mathrm{L}}$. A larger $\bm{Q}_{\mathrm{L}}$ or smaller $\bm{R}_{\mathrm{L}}$ increases pulling suppression but weakens noise rejection. Rapid KF estimation is also essential for rapid pulling suppression; this favors a larger $q_{\mathrm{p}}$ or a smaller observation covariance $\bm{R}$. After the pulling has been suppressed, the estimation speed is gradually reduced to improve Doppler-frequency estimation accuracy. On this basis, the following time-varying parameter strategy is adopted:\\
(1) $\bm{R}_{\mathrm{L}}$ is held fixed while $\bm{Q}_{\mathrm{L}}$ and $q_{\mathrm{p}}$ are varied. Initial values, $\bm{Q}_{\mathrm{L\_Init}}$ and $q_{\mathrm{p\_Init}}$, are chosen from the required pulling-convergence time. Final values, $\bm{Q}_{\mathrm{L\_Final}}$ and $\bm{q}_{\mathrm{p\_}\mathrm{Final}}$, are then chosen from the laser-noise and pulling requirements. $\bm{Q}_{\mathrm{L}}$ and $q_{\mathrm{p}}$ are varied exponentially from their initial to final values according to:
\begin{equation*}
\begin{aligned}\bm{Q}_{\mathrm L}(t)&=\bm{Q}_{\mathrm{L\_Init}}\exp\!\left(-\frac{t}{T_{\mathrm L}}\ln\frac{\bm{Q}_{\mathrm{L\_Init}}}{\bm{Q}_{\mathrm{L\_Final}}}\right)\\q_{\mathrm p}(t)&=q_{\mathrm{p\_Init}}\exp\!\left(-\frac{t}{T_{\mathrm K}}\ln\frac{q_{\mathrm{p\_Init}}}{q_{\mathrm{p\_Final}}}\right)\end{aligned}
\tag{68}\label{eq:68}
\end{equation*}
Here $T_{\mathrm{L}}$ and $T_{\mathrm{K}}$ is the duration of the LQR and KF parameter transition, respectively, and depends on the laser-noise type, controller gain, and related parameters.\\
(2) At the end of the initial stage, frequency-pulling has already been effectively suppressed. The LQR feedforward branch is therefore removed by setting all elements of $\bm{K}_{4}$ to zero, which improves the noise-suppression performance.

\subsection{Suppressor deployment}

Section II developed the suppressor model and derived explicit Doppler measurement expressions using a PI arm-locking controller. To deploy the suppressor with non-PI arm-locking controllers, we further analyze the model. In Fig.~\ref{fig:2}, if the measurement is taken at the arm-locking controller output (point $\mathrm{P}$), the transfer function from $v_{\mathrm{D}}$ to $v_{\mathrm{P}}$ is:
\begin{equation*}
H_{\mathrm P}(s)=\frac{G(s)}{1+P_{\mathrm s}(s)G(s)}=\frac{1}{1/G(s)+P_{\mathrm s}(s)}
\tag{69}\label{eq:69}
\end{equation*}
Within the science band, $1/G(s)$ is approximately zero, and in the low-frequency range, $P_\mathrm{s}(s)$ can be approximated as $s\tau$, so Eq.~\eqref{eq:68} reduces to:
\begin{equation*}
H_{P}(s)\approx\frac{1}{\tau}\frac{1}{s}
\tag{70}\label{eq:70}
\end{equation*}
Thus the arm-locking can be approximated by an integrator with coefficient, 1/$\tau$. This equivalent integrator can replace the explicit integrator in Fig.~\ref{fig:2}, with the coefficient changed to $k$=1/$\tau$.

\begin{figure}[H]
\centering
\includegraphics[width=0.50\textwidth]{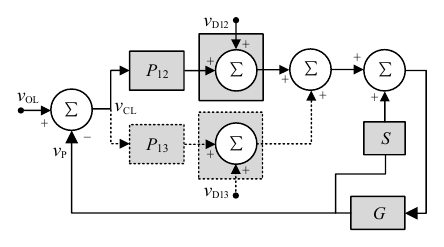}
\caption{Deployment of the suppressor in single-arm- and common-arm-locking. The dashed branch is the second arm used in common-arm-locking; removing it gives the single-arm-locking configuration.}
\label{fig:16}
\end{figure}

The suppressor can be deployed in single-arm, common-arm, dual-arm, and modified-dual-arm configurations. Fig.~\ref{fig:16} shows its deployment in the single-arm- and common-arm-locking. The blocks $P_{\mathrm{1x}}$, $G$, and $S$ denote the interferometer arm, controller, and suppressor, respectively. The Doppler-frequency $v_{\mathrm{D1}x}$ enters the arm-locking loop through the phasemeter; the suppressor measures $v_{\mathrm{D1}x}$ at the controller output, reverses its sign, and reinjects it into the loop to cancel $v_{\mathrm{D1}x}$ and suppress the resulting frequency-pulling.

\begin{figure}[!htbp]
\centering
\includegraphics[width=0.675\textwidth]{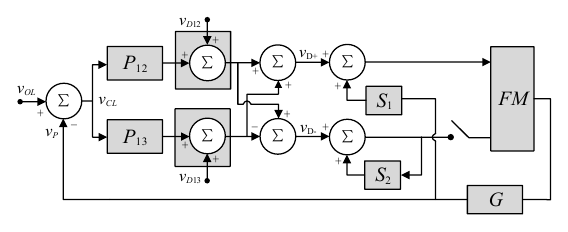}
\caption{Deployment of the suppressor in dual-arm- and modified-dual-arm-locking.}
\label{fig:17}
\end{figure}

Fig.~\ref{fig:17} shows deployment in dual-arm-locking and modified-dual-arm-locking. FM denotes the filter module, whose transfer functions for the two configurations are:
\begin{equation*}
H_{\mathrm{Dual}}(s)=v_{\mathrm{D+}}-v_{\mathrm{D-}}\frac{E(s)}{s\Delta\tau}
\tag{71}\label{eq:71}
\end{equation*}
\begin{equation*}
H_{\mathrm{MD}}(s)=v_{\mathrm{D+}}F_{\mathrm C}(s)+\left(v_{\mathrm{D+}}-v_{\mathrm{D-}}\frac{E(s)}{s\Delta\tau}\right)F_{\mathrm D}(s)
\tag{72}\label{eq:72}
\end{equation*}
Here $E(s)$ is the low-pass filter in the differential channel, $F_{\mathrm{C}}(s)$ is the low-pass filter in the common-mode channel, and $F_{\mathrm{D}}(s)$ is the high-pass filter in the dual-arm channel. $S_{1}$ and $S_{2}$ are the suppressors in the common-mode and differential-mode channels, respectively. The Doppler-frequencies $v_{\mathrm{D1}x}$ enter the loop through the phasemeters and form the common-mode signal ($v_{\mathrm{D}+}$) and differential-mode signal ($v_{\mathrm{D}-}$), which are suppressed by $S_{1}$ and $S_{2}$, respectively. After $v_{\mathrm{D}-}$ is suppressed, only $v_{\mathrm{D}+}$ passes through FM, so $v_{\mathrm{D}+}$ remains measurable. Because the differential channel contains an integrator, an unsuppressed $v_{\mathrm{D}-}$ would integrate into a large signal and severely disturb the system. A switch is therefore inserted in the differential channel. It is opened after loop closure to prevent $v_{\mathrm{D}-}$ from entering the system and is closed only after $v_{\mathrm{D}-}$ has been suppressed close to zero, thereby forming the complete dual-arm-locking.

\section{Simulation and results}

\subsection{Simulation scheme}

\begin{figure}[H]
\centering
\begin{minipage}[t]{0.46\textwidth}
\centering
\includegraphics[width=\linewidth]{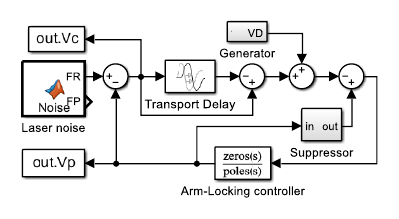}
\par\smallskip (a)
\end{minipage}
\hfill
\begin{minipage}[t]{0.53\textwidth}
\centering
\includegraphics[width=\linewidth]{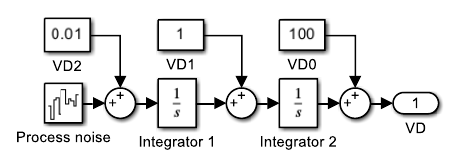}
\par\smallskip (b)
\end{minipage}
\par\medskip
\begin{minipage}[t]{0.48\textwidth}
\centering
\includegraphics[width=\linewidth]{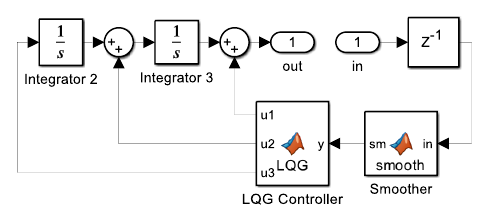}
\par\smallskip (c)
\end{minipage}
\caption{Single-arm-locking simulation diagram. (a) Arm-locking, out.Vc gives the closed-loop noise and out.Vp gives the frequency-pulling; (b) Doppler-frequency generator, $v_{\mathrm{D0}}$-$v_{\mathrm{D2}}$ is the Doppler-frequency components; (c) Suppressor, a one-step delay is inserted at the input to avoid an algebraic-loop error in the simulation.}
\label{fig:18}
\end{figure}

The single-arm-locking simulation configuration is shown in Fig.~\ref{fig:18}. In Fig.~\ref{fig:18}(a), a Transport Delay module and an arm-locking controller emulate the single-arm-locking. A Doppler-frequency generator, a Doppler-frequency suppressor, and a laser-noise module are then added to the system to complete the simulation. Fig.~\ref{fig:18}(b) shows the Doppler-frequency generator, composed of two integrators, three adders, and process noise representing higher-order terms. Fig.~\ref{fig:18}(c) shows the suppressor, including the LQG controller, measurement smoother, and Doppler-frequency simulator. The laser-noise module generates two types of laser noise: free-running (FR) and Fabry-Perot cavity-stabilized (FP), whose amplitude spectral densities (ASDs) are:
\begin{equation*}
\begin{aligned}
 \tilde v_{\mathrm{FR}}(f)&=30000\times\frac{1\,\mathrm{Hz}}{f}\;\frac{\mathrm{Hz}}{\sqrt{\mathrm{Hz}}}\\
 \tilde v_{\mathrm{FP}}(f)&=30\times\left[1+\left(\frac{2.8\,\mathrm{mHz}}{f}\right)^2\right]\frac{\mathrm{Hz}}{\sqrt{\mathrm{Hz}}}
\end{aligned}
\tag{73}\label{eq:73}
\end{equation*}

The key simulation parameters are listed in Table~\ref{tab:I}. The Transport Delay is set to 16.6 s, representative of the new LISA arm round-trip light time \cite{ghosh2022}. To balance laser-noise suppression against pulling-convergence time, the PI parameters are $g$=100 and $a$=10 or 100. The high-pass filter uses $p_{\mathrm{a}}=p_{\mathrm{b}}$=4$\times$10$^{-4}$ rad/s, corresponding to a cutoff near the lower edge of the science band (0.1 mHz). In the Doppler-frequency generator, the maximum values of $v_{\mathrm{D0}}$, $v_{\mathrm{D1}}$, and $v_{\mathrm{D2}}$ are 2.5$\times$10$^{7}$ Hz, 5 Hz/s, and 2 $\mu$Hz/$s^{2}$, respectively, and the maximum process-noise intensity is about 3$\times$10$^{-13}$. In the KF, to accelerate initial estimation, $q_{\mathrm{p\_Init}}$=1$\times$10$^{-4}$; to improve steady-state accuracy, $q_{\mathrm{p\_Final}}$=1$\times$10$^{-24}$, for the FR laser, $\bm{R}$ is approximately 1.1$\times$10$^{6}$. In the LQR, $\bm{R}_{\mathrm{L}}$=diag(1, 10, 100)$\times$10$^{2}$, $\bm{Q}_{\mathrm{L\_Init}}$ =diag(100, 10, 1) $\times$10$^{-4}$ for rapid initial pulling suppression, and $\bm{Q}_{\mathrm{L\_Final}}$ =diag(100, 10, 1) $\times$10$^{-20}$ for strong steady-state noise suppression.To acquire sufficient data for spectral analysis during the time-varying process, the LQG-parameter time-varying process is extended to 2 days.

\begin{table}[H]
\caption{Key simulation parameters.}
\label{tab:I}
\centering
\small
\setlength{\tabcolsep}{5pt}
\renewcommand{\arraystretch}{1.18}
\begin{tabular*}{0.78\textwidth}{@{\extracolsep{\fill}}cc@{}}
\hline\hline
\textbf{Module} & \textbf{Parameter and Value} \\
\hline
Transport Delay & $\tau=16.6\,\mathrm{s},\ \bar{\tau}=16.6\,\mathrm{s},\ \Delta\tau=62\,\mathrm{ms}$\\
PI Controller & $g=100,\ a=10\ \text{or}\ 100$ \\
High-pass filter & $p_{\mathrm a}=p_{\mathrm b}=4\times10^{-4}\,\mathrm{rad/s}$ \\
KF & $R=1.1\times10^{6}$ \\
 & $q_{\mathrm{p\_Init}}=1\times10^{-4},\ q_{\mathrm{p\_Final}}=1\times10^{-24}$ \\
LQR & $\bm{R}_{\mathrm L}=\operatorname{diag}(1,10,100)\times10^{2}$ \\
 & $\bm{Q}_{\mathrm{L\_Init}}=\operatorname{diag}(100,10,1)\times10^{-4}$ \\
 & $\bm{Q}_{\mathrm{L\_Final}}=\operatorname{diag}(100,10,1)\times10^{-20}$ \\
 Time-varying process &
$T_L=2\,\mathrm{days},\ T_K=4000\,\mathrm{s}$ \\
\hline\hline
\end{tabular*}
\end{table}

The goal of frequency noise suppression is to bring the ASD curve of
$v_{\mathrm{CL}}$ below the TDI curves, the capability curves of 1st- and 2nd-generation TDI are given by the following equations:
\begin{equation*}
\begin{aligned}
\tilde{v}_{\mathrm{TDI-1}}(f)
&=
1.7\times
\sqrt{
1+\left(\frac{2.8\,\mathrm{mHz}}{f}\right)^4
}
\ \frac{\mathrm{Hz}}{\sqrt{\mathrm{Hz}}}
\\
\tilde{v}_{\mathrm{TDI-2}}(f)
&=
282\times
\sqrt{
1+\left(\frac{2.8\,\mathrm{mHz}}{f}\right)^4
}
\ \frac{\mathrm{Hz}}{\sqrt{\mathrm{Hz}}}
\end{aligned}
\tag{74}\label{eq:74}
\end{equation*}

\subsection{Single-arm-locking}

\subsubsection{Laser-frequency-noise suppression capability}

\begin{figure}[!htbp]
\centering
\begin{minipage}[t]{0.495\textwidth}
\centering
\includegraphics[width=\linewidth]{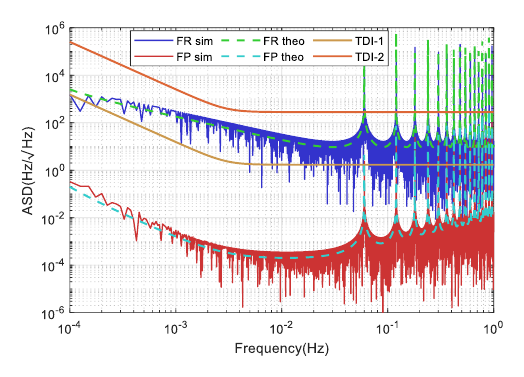}
\par\smallskip (a)
\end{minipage}
\hfill
\begin{minipage}[t]{0.495\textwidth}
\centering
\includegraphics[width=\linewidth]{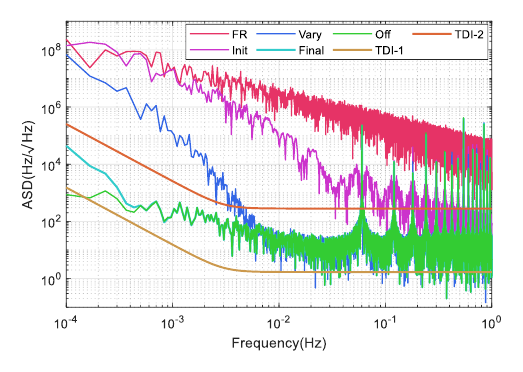}
\par\smallskip (b)
\end{minipage}
\caption{Frequency-noise suppression in the single-arm-locking for $g$=100 and $a$=100. (a) Suppressor offline: suffix 'sim' denotes simulated ASDs and 'theo' theoretical ASDs. (b) Suppressor online: FR denotes the ASD of raw Free-Running laser; Init, Vary, and Final denotes the ASDs in the initial, time-varying, and final stages, respectively; 'Off' denotes the suppressed ASD when suppressor offline.}
\label{fig:19}
\end{figure}

Laser-frequency-noise suppression is the central function of arm-locking. In the simulation, open-loop laser noise is used as the input and closed-loop noise as the output, and their spectra are compared. Because the open-loop noise contains very large low-frequency components, the time-domain data are first high-pass filtered and then multiplied by a Hamming window to reduce leakage from very-low-frequency noise before the FFT is computed. The results are shown in Fig.~\ref{fig:19}.

Fig.~\ref{fig:19}(a) shows noise suppression with the suppressor offline, so the performance is determined mainly by the PI controller and high-pass filter. Across the science band, the simulated ASDs nearly overlap the theoretical curves, validating the correctness of simulation. With $\tau$=16.6 s, the first spectral null is near 60 mHz. Under the configuration of key parameters $g$=100 and $a$=100, the suppressed FP frequency-noise simultaneously satisfied the requirements of both TDI-1 and TDI-2. In contrast, the FR frequency-noise has a relatively large amplitude and, after suppression, only satisfied the TDI-2 requirement.

Fig.~\ref{fig:19}(b) shows noise suppression with the suppressor online, for which the performance depends on both the PI controller and suppressor. FR noise is used as the input and the spectra are evaluated at different stages. During the initial stage, the ASD curve of the suppressed FP frequency-noise fails to meet the TDI-2 requirement. During the time-varying stage, only a narrow frequency interval satisfied the TDI-2 requirement. In the final stage, the ASD curve fully satisfied the TDI-2 requirement, but still not satisfied the TDI-1 requirement.

\subsubsection{Frequency-pulling suppression}

\begin{figure}[!htbp]
\centering
\begin{minipage}[t]{0.51\textwidth}
\centering
\includegraphics[width=\linewidth]{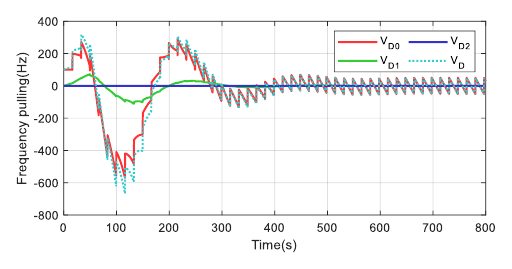}
\par\smallskip (a)
\end{minipage}
\hfill
\begin{minipage}[t]{0.48\textwidth}
\centering
\includegraphics[width=\linewidth]{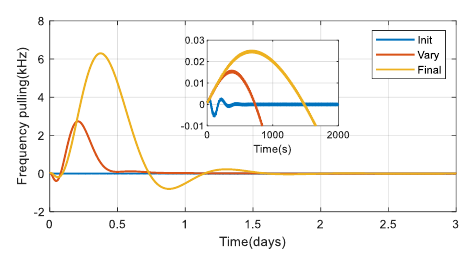}
\par\smallskip (b)
\end{minipage}
\par\medskip
\begin{minipage}[t]{0.48\textwidth}
\centering
\includegraphics[width=\linewidth]{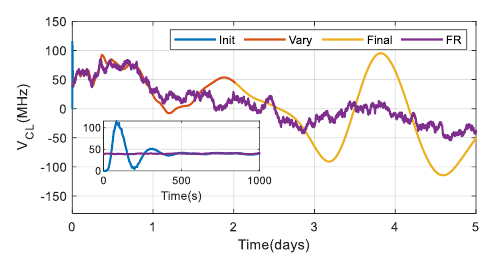}
\par\smallskip (c)
\end{minipage}
\caption{Simulation results for Frequency-pulling suppression. (a) Suppression of the three Doppler-frequency components, for $v_{\mathrm{D0}}$=100 Hz, $v_{\mathrm{D1}}$=1 Hz/s, and $v_{\mathrm{D2}}$=1 $\mu$Hz/s$^{2}$. (b) Unit-step response of $v_{\mathrm{D0}}$ in the three operating stages. (c) Open-loop and Closed-loop frequency-noise waveform during the time-varying process.}
\label{fig:20}
\end{figure}

Doppler-frequency suppression is the main focus of this work. Fig.~\ref{fig:20} summarizes several simulations. Fig.~\ref{fig:20}(a) shows the pulling generated when $v_{\mathrm{D0}}$, $v_{\mathrm{D1}}$, and $v_{\mathrm{D2}}$ act separately and together. Each pulling curve oscillates while its envelope decreases, all cases converge close to zero within approximately 500 s, demonstrating rapid pulling suppression.

Fig.~\ref{fig:20}(b) compares suppression speed in the three stages using the unit-step response of $v_{\mathrm{D0}}$. The step response tends to zero in every stage, confirming effective pulling suppression. The initial stage gives the smallest amplitude and shortest settling time. Both increase gradually during the time-varying stage and are largest in the final stage. Because the Doppler-frequency caused by the spacecraft orbit varies slowly on an annual time scale, even a final stage step-settling time of several days remains fast enough to suppress pulling driven by orbital changes.

Fig.~\ref{fig:20}(c) shows the closed-loop noise ($v_{\mathrm{CL}}$) during the time-varying process. During the initial stage, noise suppression is weak and $v_{\mathrm{CL}}$ nearly follows the FR waveform. During the time-varying stage, suppression increases progressively and the waveform evolves from strongly fluctuating to slowly varying. In the final stage, the waveform becomes smooth, showing that the high-frequency noise components have been strongly attenuated. Although the time-domain amplitude remains relatively large, its spectrum lies below the science-band requirement.

\subsection{Modified dual-arm-locking}

\begin{figure}[H]
\centering
\begin{minipage}[t]{0.57\textwidth}
\centering
\includegraphics[width=\linewidth]{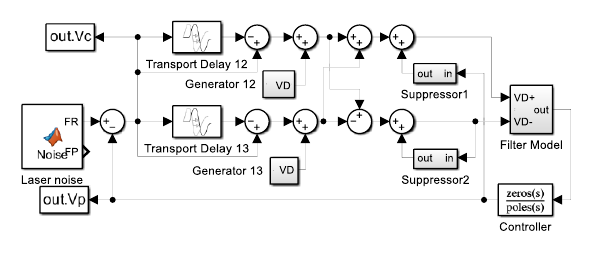}
\par\smallskip (a)
\end{minipage}
\hfill
\begin{minipage}[t]{0.42\textwidth}
\centering
\includegraphics[width=\linewidth]{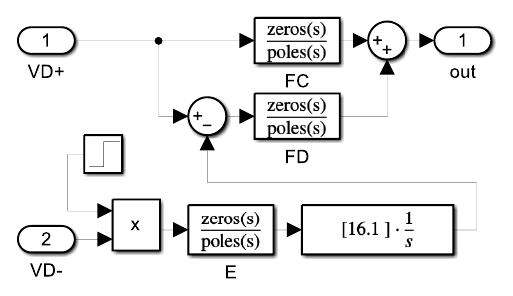}
\par\smallskip (b)
\end{minipage}
\caption{Simulation model for modified dual-arm-locking. (a) Arm-locking with $g$=100 and $a$=10; (b) Filter module, including a timed switch formed by a step signal and a multiplier.}
\label{fig:21}
\end{figure}

Because modified-dual-arm-locking combines common-arm- and dual-arm- locking, separate simulations of the common-arm and conventional dual-arm cases are omitted for brevity. The modified-dual-arm-locking simulation shown in Fig.~\ref{fig:21} adds a second arm, a differential-channel suppressor, a differential-channel suppressor, and a filter module to the single-arm-locking model. The filter module contains $E$, $F_{\mathrm{C}}$, and $F_{\mathrm{D}}$, whose transfer functions are:
\begin{equation*}
E(s)=g_1\frac{1}{s(s+p_1)},\quad F_{\mathrm C}(s)=g_2g_3\frac{s+z_3}{s(s+p_3)},\quad F_{\mathrm D}(s)=g_4g_5g_6\frac{s^4}{(s+p_4)(s+p_5)(s+p_6)^2}
\tag{75}\label{eq:75}
\end{equation*}

The filter parameters are listed in Table~\ref{tab:II}. To avoid an unstable transient immediately after loop closure, the differential channel is enabled only after 2000 s.

\begin{table}[H]
\caption{Filter parameters.}
\label{tab:II}
\centering
\small
\setlength{\tabcolsep}{14pt}
\renewcommand{\arraystretch}{1.22}
\begin{tabular*}{0.92\textwidth}{@{\extracolsep{\fill}}cccc@{}}
\hline\hline
\textbf{Filter} & \textbf{Zeros (rad/s)} & \textbf{Poles (rad/s)} & \textbf{Gain} \\
\hline
$E$ &  & $p_1=\pi$ & $g_1=\pi$ \\
$F_{\mathrm C}$ &  & $p_2=0$ & $g_2=1/\bar\tau$ \\
 & $z_3=2\pi\times5/(13\bar\tau)$ & $p_3=2\pi\times5/(2\bar\tau)$ & $g_3=p_3/z_3$ \\
$F_{\mathrm D}$ & $z_4=0$ & $p_4=7/(5\bar\tau)$ & $g_4=1$ \\
 & $z_5=0$ & $p_5=11/(20\bar\tau)$ & $g_5=1$ \\
 & $z_6=0$ & $p_6=2\pi\times1/(90\bar\tau)$ & $g_6=1$ \\
\hline\hline
\end{tabular*}
\end{table}

\begin{figure}[H]
\centering
\begin{minipage}[t]{0.495\textwidth}
\centering
\includegraphics[width=\linewidth]{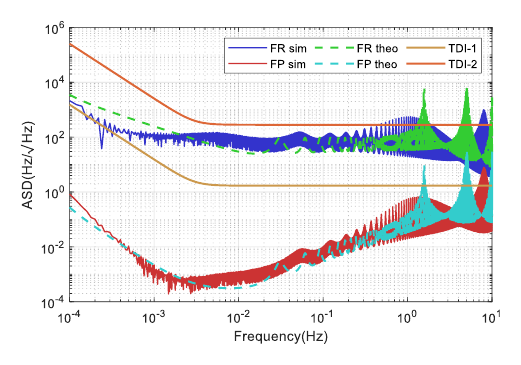}
\par\smallskip (a)
\end{minipage}
\hfill
\begin{minipage}[t]{0.495\textwidth}
\centering
\includegraphics[width=\linewidth]{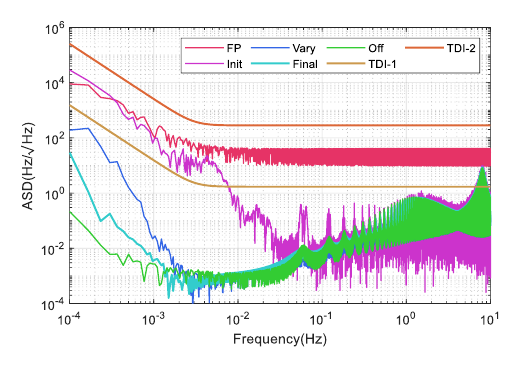}
\par\smallskip (b)
\end{minipage}
\par\medskip
\begin{minipage}[t]{0.495\textwidth}
\centering
\includegraphics[width=\linewidth]{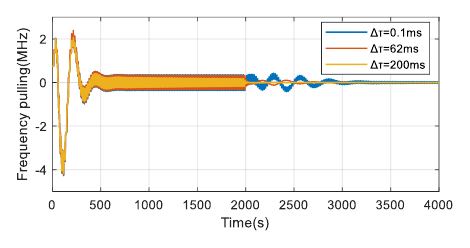}
\par\smallskip (c)
\end{minipage}
\hfill
\begin{minipage}[t]{0.495\textwidth}
\centering
\includegraphics[width=\linewidth]{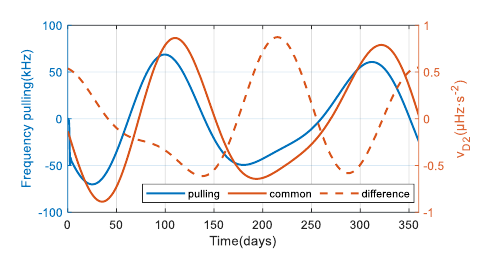}
\par\smallskip (d)
\end{minipage}
\caption{Simulation results for modified-dual-arm-locking. (a) ASDs with the suppressor offline; (b) ASDs with the suppressor online for MZ laser noise; (c) Frequency-pulling transient after the differential channel is enabled; (d) Frequency-pulling in steady-state, the common- and difference-signal curves of $v_{\mathrm{D2}}$ from Ref. \cite{mckenzie2009}.}
\label{fig:22}
\end{figure}

The simulation results are shown in Fig.~\ref{fig:22}. Fig.~\ref{fig:22}(a) gives the noise-suppression performance with the suppressor offline. The ASDs of all two suppressed frequency-noise satisfied the minimum TDI-2 requirement. Unlike single-arm-locking, the first sensor null is shifted to 5 Hz and therefore lies above the full science band. Fig.~\ref{fig:22}(b) shows the suppressor-online case and using an FP laser as input. Because the ASD of the raw FP-noise already satisfied the TDI-2 requirement, in the initial stage, the ASD curve satisfied the TDI-2 requirement but not the TDI-1 requirement. In the time-varying stage, noise suppression improves continuously with operating time, in the final stage, the ASD curve eventually satisfied the TDI-1 requirement. Fig.~\ref{fig:22}(c) shows the transient when the differential channel is enabled. The configuration changes from common-arm-locking to modified dual-arm-locking, the first null moves to 5 Hz and the 60 mHz oscillatory component is filtered, making the waveform smoother. In addition, for different values of $\Delta\tau$, the curves change only slightly; the only difference is that the convergence time is slightly longer when $\Delta\tau$ = 0.1 ms. Fig.~\ref{fig:22}(d) shows the variation curves of frequency pulling over one year in the steady-state under common-mode and differential-mode signal driving of $v_{\mathrm{D2}}$, and the maximum frequency-pulling amplitude is only approximately 80 kHz.

\subsection{Combined with non-PI arm-locking controllers}

\begin{figure}[!htbp]
\centering
\begin{minipage}[t]{0.49\textwidth}
\centering
\includegraphics[width=\linewidth]{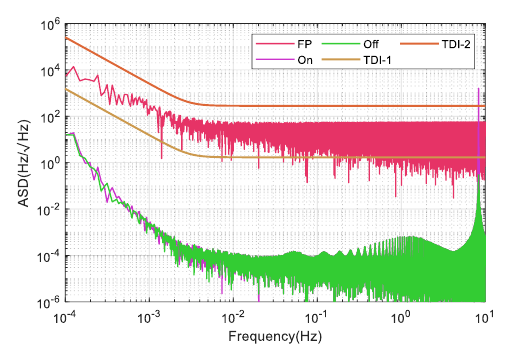}
\par\smallskip (a)
\end{minipage}
\hfill
\begin{minipage}[t]{0.495\textwidth}
\centering
\includegraphics[width=\linewidth]{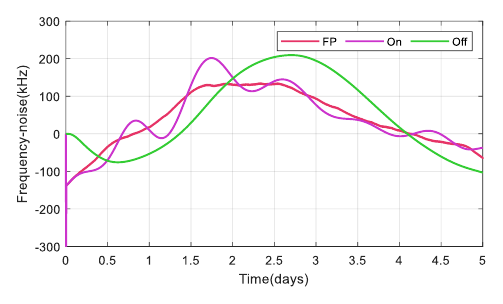}
\par\smallskip (b)
\end{minipage}
\par\medskip
\begin{minipage}[t]{0.50\textwidth}
\centering
\includegraphics[width=\linewidth]{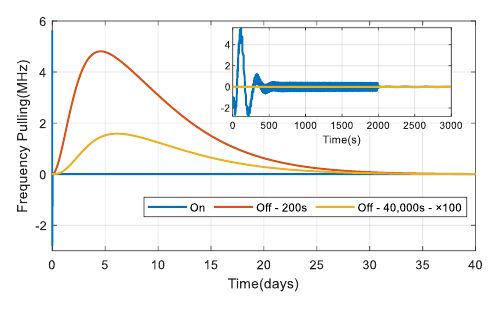}
\par\smallskip (c)
\end{minipage}
\caption{Simulation of the suppressor combined with other arm-locking controller. (a) FP laser frequency-noise suppression. (b) Frequency-noise curves. ‘FP' denotes the raw noise of the FP laser; 'On' and 'Off' denote the frequency noise with the suppressor online and offline, respectively, and the Doppler frequency estimation error is set to zero; (c)Frequency pulling, “Off - 200s” denotes the case in which the suppressor is offline and the Doppler frequency estimation average time is 200 s; “Off - 40,000s - ×100” denotes the case in which the suppressor is offline, the average time is 40,000 s, and the curve is magnified by a factor of 100 for ease of comparison. Simulation parameters are listed in Table~\ref{tab:III}.}
\label{fig:23}
\end{figure}

The suppressor is also added to the arm-locking of Ref. \cite{ghosh2022} to verify compatibility with other controllers than the PI controller used above. Unlike the other simulations, the time-varying process is shortened to 7200 s, the results are shown in Fig.~\ref{fig:23}. Fig.~\ref{fig:23}(a) shows the suppression of FP laser frequency-noise. Similar to the case with a PI arm-locking controller, in the final stage, the frequency-noise after suppression satisfied the requirements of both TDI-1 and TDI-2, and the curve is almost the same as that obtained with the suppressor offline. Fig.~\ref{fig:23}(b) presents the frequency noise of the FP laser, with the suppressor online and offline. The small differences in the maximum amplitudes of the three curves indicate that shortening the time-varying process to 7200 s does not induce significant frequency pulling. Fig.~\ref{fig:23}(c) compares pulling with the suppressor offline and online. With the suppressor online, the maximum pulling is about 5 MHz and the convergence time is only about 2000 s. With the suppressor offline, the maximum pulling is about 5 MHz and convergence requires more than 30 days.Simulation results show that after the suppressor is applied to the arm-locking controller, the frequency pulling is rapidly suppressed. Fig.~\ref{fig:23}(c) compares the frequency pulling with the suppressor online and offline. When the suppressor is online, the maximum amplitude of the frequency pulling is approximately 5 MHz, and the convergence time is approximately 2000 s. When the suppressor is offline, Doppler frequency estimation is performed using averaging times of 200 s and 40,000 s, respectively. The frequency pulling caused by the estimation error is approximately 5 MHz and 16 kHz, respectively. Taking convergence of the frequency pulling to 10 kHz as the criterion, the convergence times are approximately 30 days and 10 days, respectively.

\begin{table}[H]
\caption{Parameters used in the frequency-pulling simulation.
$v_{\mathrm{DE}}$ denotes Doppler-frequency estimation error,
subscripts `+' and `-' denotes common- and differential-mode quantities,
the values are taken from Ref.~[16] and [24].}
\label{tab:III}

\centering
\small
\renewcommand{\arraystretch}{1.25}

\begin{tabular*}{\linewidth}{@{\extracolsep{\fill}}ccccc}

\hline
\hline

\rule{0pt}{20pt}
\begin{tabular}{c}
Doppler\\
frequency
\end{tabular}
&
\begin{tabular}{c}
Estimation\\
average time(s)
\end{tabular}
&
\begin{tabular}{c}
Constant\\
term ($\mathrm{Hz}$)
\end{tabular}
&
\begin{tabular}{c}
1st-order term\\
($\mathrm{Hz/s}$)
\end{tabular}
&
\begin{tabular}{c}
2nd-order term\\
($\mathrm{Hz/s^2}$)
\end{tabular}
\\

\hline

$v_{\mathrm{D+}}$
&
&
$-2\times10^{6}$
&
$-4.5$
&
$-0.4\times10^{-6}$
\\[2pt]

$v_{\mathrm{D-}}$
&
&
$-7\times10^{6}$
&
$-1.3$
&
$1.1\times10^{-6}$
\\[2pt]

$v_{\mathrm{DE+}}$
&
$200$
&
$6.88$
&
$1.308\times10^{-2}$
&
$-2.4\times10^{-7}$
\\[2pt]

$v_{\mathrm{DE-}}$
&
$200$
&
$-0.04$
&
$7.71\times10^{-5}$
&
$-1.2\times10^{-7}$
\\[2pt]

$v_{\mathrm{DE+}}$
&
$40,000$
&
$4.89\times10^{-3}$
&
$1.96\times10^{-6}$
&
$8.43\times10^{-10}$
\\[2pt]

$v_{\mathrm{DE-}}$
&
$40,000$
&
$-2.89\times10^{-5}$
&
$-1.15\times10^{-8}$
&
$-4.98\times10^{-12}$
\\

\hline
\hline

\end{tabular*}

\end{table}

\section{Discussion}

This work focuses on Doppler-frequency pulling. Because pulling suppression and laser-noise suppression are inherently competing objectives, an arm-locking should emphasize different objectives at different stages. We therefore discuss these two aspects separately.

Regarding Doppler-frequency pulling, the simulation results show that the suppressor rapidly mitigates the pulling effect. During the suppression process, the pulling magnitude remains below five times the Doppler frequency. At the maximum Doppler frequency of 20 MHz \cite{mckenzie2009}, the resulting frequency pulling is approximately 100 MHz, corresponding to only 1\% of the mode-hop-free range and therefore remaining within a safe operating range. Consequently, the Doppler-frequency pre-estimation stage can be omitted. In the steady state, the suppressor retains sufficient frequency-pulling suppression capability to accommodate pulling caused by variations in the spacecraft orbits, with the residual pulling magnitude remaining below 100 kHz.

Regarding laser frequency-noise suppression, the arm-locking exhibits relatively weak suppression immediately after loop closure. However, its suppression performance gradually improves with operating time and eventually reaches the level achieved when the suppressor is operated offline. The noise-suppression performance can be adjusted through the LQG parameters. The analysis demonstrates that the arm-locking remains stable, allowing a high level of noise suppression to be imposed. The associated cost, however, is a longer time-varying process. These two requirements should therefore be carefully balanced when selecting the controller parameters.

A comparison between the offline and online implementations of the suppressor reveals distinct performance characteristics. In the offline case, the maximum frequency pulling depends on the Doppler-frequency estimation error and the arm-locking controller parameters: larger estimation errors and higher closed-loop gains result in stronger frequency pulling. In the online case, the maximum frequency pulling is governed primarily by the constant component of the Doppler frequency at loop closure, while its peak magnitude remains below five times this constant component. Furthermore, the lock-acquisition time in the offline case depends on both the Doppler-frequency estimation error and the convergence rate of the frequency pulling. By contrast, in the online case, it is determined mainly by the time at which the ASD curve of $v_{\mathrm{CL}}$ falls below the TDI curve. The simulation results in Fig.~\ref{fig:23} are used to compare the time required for the suppressor in the offline and online cases, and the corresponding data are listed in Table~\ref{tab:IV}. Using the total time spent in all stages as the basis for comparison, the data show that less time is required when the suppressor is online.
\begin{table}[H]
\caption{Time Required for Each Stage, where the Doppler frequency estimation is performed by three-stage averaging, so the estimation time is three times the averaging time.}
\label{tab:IV}

\centering
\small
\renewcommand{\arraystretch}{1.25}

\begin{tabular*}{\linewidth}{@{\extracolsep{\fill}}ccccc}

\hline
\hline

\begin{tabular}{c}
Suppressor
\end{tabular}
&
\begin{tabular}{c}
Doppler-frequency\\
estimation time
\end{tabular}
&
\begin{tabular}{c}
Pulling convergence\\
time
\end{tabular}
&
\begin{tabular}{c}
Time-varying time
\end{tabular}
&
\begin{tabular}{c}
Total time
\end{tabular}
\\

\hline

Online
&
$0\,\mathrm{s}$
&
$2000\,\mathrm{s}$
&
$7200\,\mathrm{s}$
&
$9200\,\mathrm{s}$
\\[3pt]

Offline
&
$600\,\mathrm{s}$
&
$\approx30\,\mathrm{days}$
&
$0\,\mathrm{s}$
&
$\approx30\,\mathrm{days}$
\\[3pt]

offline
&
$120000\,\mathrm{s}$
&
$\approx10\,\mathrm{days}$
&
$0\,\mathrm{s}$
&
$\approx12\,\mathrm{days}$
\\

\hline
\hline

\end{tabular*}

\end{table}
After the suppressor is introduced, the acquisition time depends strongly on the duration of the LQG parameter transition. The present exponential decay schedule may not be optimal. Improving KF estimation accuracy is also a direct route to lower low-frequency residual noise, but the present linear filtering model is not necessarily optimal; adaptive estimation may provide further improvement. Optimizing both the parameter schedule and the filtering model may therefore shorten acquisition further.

\section{Conclusion}

This work has presented a real-time Doppler-frequency estimation and suppression method through system modeling, suppressor design, and a time-varying LQG parameter strategy. Comprehensive simulations in single-arm and modified-dual-arm systems verify the feasibility of the suppressor. During pulling convergence, the maximum frequency-pulling amplitude remains below 100 MHz and the convergence time is shorter than 1000 s, allowing the Doppler pre-estimation stage to be omitted. During the time-varying LQG process, the closed-loop laser noise decreases rapidly to satisfy the TDI requirement and ultimately approaches the suppression performance obtained with the suppressor offline. In the modified-dual-arm system, the suppressor substantially reduces the pulling amplitude, pulling duration, and overall lock-acquisition time, demonstrating its practical value.

Further reductions in acquisition time and improvements in observing efficiency may be achieved by optimizing the time-varying parameter schedule and improving Doppler-frequency estimation accuracy. Future work will model different types of laser noise and develop more efficient parameter-variation strategies to improve the overall suppressor performance.

\begin{acknowledgments}
This work is supported by the National Key R\&D Program of China (Grant No. 2022YFC2204602), the National Natural Science Foundation of China (Grant Nos. 12505072, 1227050027 and 12150012), the China Postdoctoral Science Foundation-Hubei Joint Support Program (Grant No. 2025T005HB), the China Postdoctoral Science Foundation (Grant Nos. 2024M760994 and GZB20250771).
\end{acknowledgments}

\appendix
\section{Step response of the Doppler frequency}

Because the arm-locking transfer function contains a delay, it is transcendental and its step response is cumbersome to analyze directly in the frequency domain. We therefore analyze the response in the time domain. For the single-arm-locking in Fig.~\ref{fig:2}, during the first light-propagation cycle after loop closure, $v_{\mathrm{D0}}$ enters through the phasemeter as a step, denoted $v_{i\_1}$. The signals generated at points A and CL are denoted $v_{\mathrm{A0\_1}}$ and $v_{\mathrm{CL\_1}}$, respectively:
\begin{equation*}
v_{\mathrm{i\_1}}(s)=\frac{v_{\mathrm{D0}}}{s},\qquad v_{\mathrm{A0\_1}}(s)=v_{\mathrm{i\_1}}\frac{1}{1+G(s)},\qquad v_{\mathrm{CL\_1}}(s)=v_{\mathrm{i\_1}}\frac{-G(s)}{1+G(s)}
\tag{A1}\label{eq:A1}
\end{equation*}
The signal $v_{\mathrm{CL\_1}}$ simultaneously propagates through the interferometer arm.

At the start of the second cycle, the subscripts are incremented. The propagated $v_{\mathrm{CL\_1}}$ signal arrives at point B with reversed sign and becomes the input to the second cycle. The corresponding signals are
\begin{equation*}
v_{\mathrm{i\_2}}(s)=-v_{\mathrm{CL\_1}}(s),\qquad v_{\mathrm{A0\_2}}(s)=v_{\mathrm{i\_2}}\frac{1}{1+G(s)},\qquad v_{\mathrm{CL\_2}}(s)=v_{\mathrm{i\_2}}\frac{-G(s)}{1+G(s)}
\tag{A2}\label{eq:A2}
\end{equation*}
Proceeding recursively, the signals in the nth cycle are
\begin{equation*}
v_{\mathrm{i\_n}}(s)=-v_{\mathrm{CL\_n-1}}(s),\qquad v_{\mathrm{A0\_n}}(s)=v_{\mathrm{i\_n}}\frac{1}{1+G(s)},\qquad v_{\mathrm{CL\_n}}(s)=v_{\mathrm{i\_n}}\frac{-G(s)}{1+G(s)}
\tag{A3}\label{eq:A3}
\end{equation*}
By induction, the nth-cycle expression for $v_{\mathrm{A0\_n}}$ is
\begin{equation*}
v_{\mathrm{A0\_n}}(s)=\frac{v_{\mathrm{D0}}}{s}\frac{G^{n-1}(s)}{[1+G(s)]^n}
\tag{A4}\label{eq:A4}
\end{equation*}
Multiplying the signal in each propagation cycle by the corresponding delay factor and summing the cycles gives the complete response:
\begin{equation*}
v_{\mathrm{A0}}(s)=\sum_{n=1}^{\infty}v_{\mathrm{A0\_n}}(s)e^{-(n-1)\tau s}
\tag{A5}\label{eq:A5}
\end{equation*}
Substitution of the PI-controller transfer function $G(s)$ yields
\begin{equation*}
v_{\mathrm{A0}}(s)=v_{\mathrm{D0}}\sum_{n=1}^{\infty}
 \frac{\left[g(s+a)e^{-s\tau}\right]^{n-1}}{\left[(g+1)s+ga\right]^n}
\tag{A6}\label{eq:A6}
\end{equation*}

\section{Step response of laser frequency noise}

For the single-arm-locking, the derivation is analogous to Appendix A. During the first light-propagation cycle after loop closure in Fig.~\ref{fig:2}, the system input is denoted $v_{i\_1}$ and the signals at points A and CL are $v_{\mathrm{AL\_1}}$ and $v_{\mathrm{CL\_1}}$, respectively:
\begin{equation*}
v_{\mathrm{i\_1}}(s)=\frac{v_{\mathrm L}}{s},\qquad v_{\mathrm{AL\_1}}(s)=v_{\mathrm{i\_1}}\frac{1}{1+G(s)},\qquad v_{\mathrm{CL\_1}}(s)=v_{\mathrm{i\_1}}\frac{1}{1+G(s)}
\tag{B1}\label{eq:B1}
\end{equation*}
The signal $v_{\mathrm{CL\_1}}$ simultaneously propagates through the interferometer arm.

At the start of the second cycle, the subscripts are incremented. The propagated $v_{\mathrm{CL\_1}}$ signal reaches point B with reversed sign and becomes the input to the second cycle. The corresponding signals are
\begin{equation*}
v_{\mathrm{i\_2}}(s)=-v_{\mathrm{CL\_1}}(s),\qquad v_{\mathrm{AL\_2}}(s)=v_{\mathrm{i\_2}}\frac{1}{1+G(s)},\qquad v_{\mathrm{CL\_2}}(s)=v_{\mathrm{i\_2}}\frac{-G(s)}{1+G(s)}
\tag{B2}\label{eq:B2}
\end{equation*}
Proceeding recursively, the signals in the nth cycle are
\begin{equation*}
v_{\mathrm{i\_n}}(s)=-v_{\mathrm{CL\_n-1}}(s),\qquad v_{\mathrm{AL\_n}}(s)=v_{\mathrm{i\_n}}\frac{1}{1+G(s)},\qquad v_{\mathrm{CL\_n}}(s)=v_{\mathrm{i\_n}}\frac{-G(s)}{1+G(s)}
\tag{B3}\label{eq:B3}
\end{equation*}
By induction, the nth-cycle expression for $v_{\mathrm{AL\_n}}$ is
\begin{equation*}
v_{\mathrm{AL\_n}}(s)=\frac{v_{\mathrm L}}{s}\times\begin{cases}\dfrac{1}{1+G(s)},&n=1\\[7pt]\dfrac{-G^{n-2}(s)}{[1+G(s)]^n},&n>1\end{cases}
\tag{B4}\label{eq:B4}
\end{equation*}
Multiplying each propagation-cycle signal by its delay factor and summing the cycles gives the complete response:
\begin{equation*}
v_{\mathrm{AL}}(s)=\sum_{n=1}^{\infty}v_{\mathrm{AL\_n}}(s)e^{-(n-1)\tau s}
\tag{B5}\label{eq:B5}
\end{equation*}
Substitution of the PI-controller transfer function $G(s)$ yields
\begin{equation*}
v_{\mathrm{AL}}(s)=\frac{v_{\mathrm L}}{(g+1)s+ga}-v_{\mathrm L}\sum_{n=2}^{\infty}\frac{s\left[g(s+a)\right]^{n-2}e^{-(n-1)s\tau}}{\left[(g+1)s+ga\right]^n}
\tag{B6}\label{eq:B6}
\end{equation*}

\nocite{apsrev42Control}
\bibliographystyle{apsrev4-2}
\bibliography{references}

\end{document}